\documentclass[sigconf]{acmart}

\renewcommand\footnotetextcopyrightpermission[1]{} 
\graphicspath{{./figures/}}
\def\ourprotocol{{{TEMPEST-LoRa}}\xspace}

\usepackage{multirow}
\usepackage{amsmath}
\usepackage{amsfonts}
\usepackage{amssymb}
\usepackage{balance}
\usepackage{bm}
\usepackage{graphics}
\usepackage{graphicx}
\usepackage{grffile}
\usepackage{footnote}
\usepackage{pifont}
\usepackage{epsfig}
\usepackage{xspace}
\usepackage{mathtools}
\usepackage{array}
\usepackage{etoolbox}
\usepackage{enumitem}
\usepackage{subfigure}
\usepackage{booktabs}
\usepackage{tablefootnote}
\usepackage{hyperref}
\usepackage{verbatim}
\usepackage{makecell}
\usepackage{algorithm}  
\usepackage{algpseudocode}
\usepackage{listings}
\usepackage{bbding}
\usepackage{tikz}
\usepackage{amsmath}
\usepackage{nicefrac}
\usepackage{siunitx}
\usepackage{array,framed}
\usepackage{booktabs}
\usepackage{diagbox}

\newcommand{\ie}{\textit{i}.\textit{e}.}
\newcommand{\eg}{\textit{e}.\textit{g}.}
\newcommand{\etc}{\textit{etc}}

\usepackage{
  color,
  float,
  epsfig,
  wrapfig,
  graphics,
  graphicx
}

\AtBeginDocument{%
  }

\acmConference[CCS '25]{the 2025 ACM SIGSAC Conference on Computer and Communications Security (CCS '25)}{October 13--17, 2025}{Taipei, Taiwan}

\begin{document}

\title{TEMPEST-LoRa: Cross-Technology Covert Communication}

\author{Xieyang Sun}
\affiliation{%
  \institution{Xi'an Jiaotong University}
  \city{Xi'an}
  \country{China}
}
\email{xieyangsun@stu.xjtu.edu.cn}

\author{Yuanqing Zheng}
\affiliation{%
  \institution{The Hong Kong Polytechnic University}
  \city{Hong Kong}
  \country{China}
}
\email{csyqzheng@comp.polyu.edu.hk}

\author{Wei Xi}
\authornote{Wei Xi is the corresponding author.}
\affiliation{%
  \institution{Xi'an Jiaotong University}
  \city{Xi'an}
  \country{China}
}
\email{xiwei@xjtu.edu.cn}

\author{Zuhao Chen}
\affiliation{%
  \institution{Xi'an Jiaotong University}
  \city{Xi'an}
  \country{China}
}
\email{czh869452912@gmail.com}

\author{Zhizhen Chen}
\affiliation{%
  \institution{Xi'an Jiaotong University}
  \city{Xi'an}
  \country{China}
}
\email{zhizhenc@stu.xjtu.edu.cn}

\author{Han Hao}
\affiliation{%
  \institution{Xi'an Jiaotong University}
  \city{Xi'an}
  \country{China}
}
\email{haohan9717@stu.xjtu.edu.cn}

\author{Zhiping Jiang}
\affiliation{%
  \institution{Xidian University}
  \city{Xi'an}
  \country{China}
}
\email{zpj@xidian.edu.cn}

\author{Sheng Zhong}
\affiliation{%
  \institution{Nanjing University}
  \city{Nanjing}
  \country{China}
}
\email{zhongsheng@nju.edu.cn}



\begin{abstract}

Electromagnetic (EM) covert channels pose significant threats to computer and communications security in air-gapped networks. 
Previous works exploit EM radiation from various components (\eg, video cables, memory buses, CPUs) to secretly send sensitive information. 
These approaches typically require the attacker to deploy highly specialized receivers near the victim, which limits their real-world impact.  
This paper reports a new EM covert channel, TEMPEST-LoRa, that builds on Cross-Technology Covert Communication (CTCC), which could allow attackers to covertly transmit EM-modulated secret data from air-gapped networks to widely deployed operational LoRa receivers from afar.
We reveal the potential risk and demonstrate the feasibility of CTCC by tackling practical challenges involved in manipulating video cables to precisely generate the EM leakage that could readily be received by third-party commercial LoRa nodes/gateways. 
Experiment results show that attackers can reliably decode secret data modulated by the EM leakage from a video cable at a maximum distance of 87.5m or a rate of 21.6 kbps. We note that the secret data transmission can be performed with monitors turned off (therefore covertly).
\end{abstract}





\maketitle

\begingroup
\renewcommand\thefootnote{\fnsymbol{footnote}}
\footnotetext{This paper has been accepted to ACM CCS 2025. This is the authors' preprint version.}
\endgroup

\begin{figure}[t]
\centering
\includegraphics[width=0.99\linewidth]{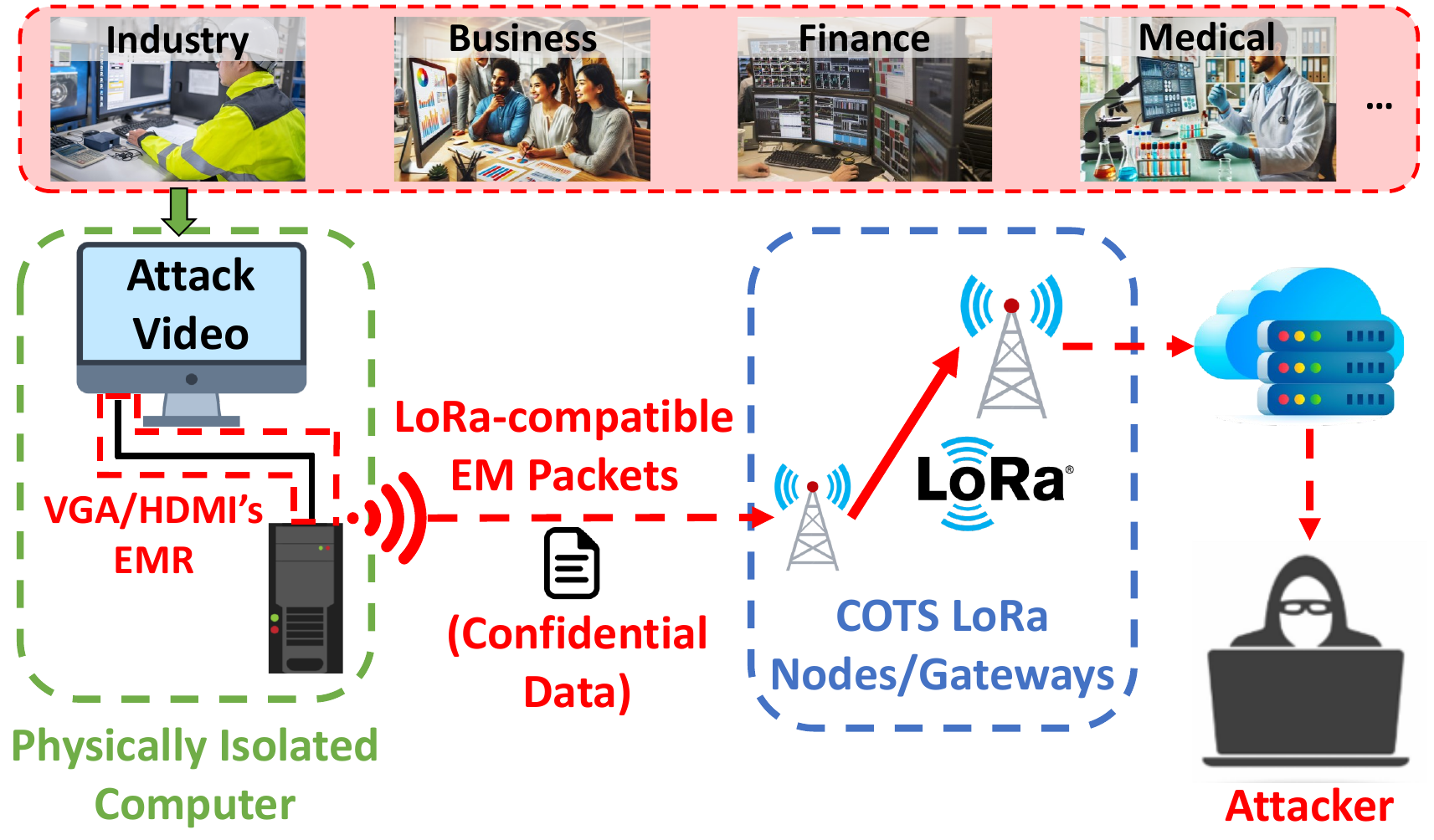}
\caption{\ourprotocol reveals the risk of a new cross-technology covert communication attack, where attackers can manipulate video cables to generate LoRa-compatible EM packets, which can be received and processed by operational LoRa nodes/gateways widely deployed worldwide.}
\label{fig:AttackTarget}
\end{figure}

\section{Introduction}
\label{sec:intro}

Physically isolated (air-gapped) networks are among the most effective ways of enhancing the computer and communications security against real-world attacks~\cite{Stuxnet,Ramsay} in industry, business, finance, and medical sectors.
Electromagnetic (EM) covert channels, however, pose serious threats to the physically isolated networks~\cite{Bridgeware,lavaud2021whispering}. Previous research has demonstrated that implanted malware can manipulate electromagnetic radiation (EMR) of computer components (such as DRAM~\cite{GSMem,EMLoRa}, USB~\cite{USBee}), thereby covertly modulating and exfiltrating confidential data to covert receivers. The EM covert channels can bypass the air-gapped systems independently of traditional communication channels (\eg, Internet, WiFi, Bluetooth, \etc). 
Constrained by the emission characteristics (\eg, ultra-low power of EMR, EM modulation fidelity and resolution, and data rate), existing covert channel attacks typically require very short communication ranges (\eg, <10m) and necessitate the physical deployment of highly specialized bulky receivers (\eg, high-end software defined radios) close to the isolated networks, which practically limit their risks and real-world implications so far. 

In this paper, we reveal a new risk of EM covert channels, which could allow attackers to exploit existing  operational LoRa (\underline{Lo}ng \underline{Ra}nge Radio~\cite{li2022lora}) nodes/gateways widely deployed worldwide to receive secret data leaked from air-gapped networks at much greater distances and higher data rates. Some third-party LoRa devices are freely accessible to attackers across the globe. To demonstrate the feasibility of such cross-technology (\ie, EMR to LoRa) covert communication (CTCC), we develop \ourprotocol, which modulates EMR from video cables (VGA or HDMI) and thereby generates EMR-modulated LoRa packets. By doing so, attackers can leverage Commercial Off-The-Shelf (COTS) LoRa gateways as receivers to covertly receive secret data as depicted in Figure~\ref{fig:AttackTarget}. Unlike previous work that necessitates close proximity from the EMR sources, the LoRa-like EMR signals can penetrate a few concrete walls and retain a sufficiently high signal strength over long propagation distances thanks to the unique noise-resilience feature of LoRa modulation and the high sensitivity of LoRa receivers~\cite{SX1262}.

We demonstrate that realizing this new type of covert communication, though challenging, is indeed possible for attackers at a very low attack launching cost and has a substantial security implication. To generate LoRa-compatible EMR signals with video cables, attackers must accurately modulate the EM leakage, which requires a new technical design. Although previous studies \cite{SoftTEMPEST,TEMPESTForEliza,AirHopper,LCDReloaded} have investigated the problem of generating the EMR by manipulating video cables, the modulation fidelity and resolution of existing works fall short in supporting the cross-technology covert communication with commercial wireless protocols such as LoRa. 

We revisit the EMR model of video cables and develop a novel fine-grained pixel-level EM modulation method. In particular, we repurpose a video cable as a direct radio-frequency (RF) sampling transmitter, enabling high-fidelity modulation up to the pixel clock (PC) frequency. With this new EMR control technology, a curated attack image can cause variations of electronic signals over the video cable to generate EMR at different frequencies. However, the PC frequency is typically much lower than the wireless frequency of LoRa gateways (\eg, 915 MHz in US). To address this problem, we exploit the harmonics of EMR and shift the frequencies to the target LoRa band, making the EM packets readily decodable by operational LoRa nodes/gateways, widely deployed and freely accessible to attackers worldwide. 

To ensure covertness during the secret data transmission, we also study how to achieve visual invisibility on the victim's monitor. We find that by modifying the power management interface of the monitor using DDCcontrol~\cite{DDCcontrol}, attackers can deactivate the monitors while keeping the video cables active and continuously emitting EM packets. This could allow attackers to covertly send the secret data with a black display on the screen.

This paper makes the following key contributions: 
\begin{itemize}
    \item We develop a fine-grained pixel-level EMR modulation technique that transforms a video cable into a direct RF sampling transmitter, enabling attackers to generate protocol-compatible EMR signals for cross-technology covert communication with wireless protocols such as LoRa. 
    \item We reveal the risk of a new type of covert channel from EMR to operational LoRa devices. Unlike previous works, this new covert channel poses a unique threat to air-gapped networks, since LoRa-compatible EM packets can penetrate thick concrete-walls over long attacking ranges and be directly received by LoRa nodes/gateways deployed worldwide.
    \item We prototype \ourprotocol and evaluate the performance in various practical settings.  
    Our experiment results with both VGA and HDMI cables show that \ourprotocol can covertly transmit secret data to COTS LoRa nodes or gateways at a maximum rate of 21.6 bps or up to 87.5m away, and even further to low-cost software-defined radios (SDR) such as HackRF One with customized EM packets. 
\end{itemize}

\section{Related Work}

\begin{table}[t]
\caption{\ourprotocol v.s. other works.}
\resizebox{0.99\columnwidth}{!}
{
\begin{tabular}{c|c|c|c|c}
\toprule[0.5mm]
\textbf{Method} & \begin{tabular}[c]{@{}c@{}}\textbf{Leak}\\ \textbf{source}\end{tabular} & \textbf{Range} & \textbf{Speed} & \textbf{Receiver} \\ \hline
GSMem~\cite{GSMem} & DRAM & 1-5.5m & \begin{tabular}[c]{@{}c@{}}100-1kbps\end{tabular} & \begin{tabular}[c]{@{}c@{}}Phone w/\\modified firmware\end{tabular} \\ \hline
BitJabber~\cite{Bitjabber} & DRAM & \textless{}3m & \begin{tabular}[c]{@{}c@{}}100k-300kbps\end{tabular} & SDR \\ \hline
NoiseSDR~\cite{Noise-SDR} & DRAM & \textless{}5m & \begin{tabular}[c]{@{}c@{}}11.2-2.56kbps\end{tabular} & \begin{tabular}[c]{@{}c@{}}SDR\end{tabular} \\ \hline
Air-Fi~\cite{Air-Fi} & DRAM & 2.1-8m & \begin{tabular}[c]{@{}c@{}}1-16bps\end{tabular} & \begin{tabular}[c]{@{}c@{}}WiFi w/\\modified firmware\end{tabular} \\ \hline
\begin{tabular}[c]{@{}c@{}}TEMPEST\\ for Eliza~\cite{TEMPESTForEliza} \end{tabular} & \begin{tabular}[c]{@{}c@{}}Video\\ Cable\end{tabular} & \textless{}5m & unknown & AM radio \\  \hline
AirHopper~\cite{AirHopper} & \begin{tabular}[c]{@{}c@{}}Video Cable\end{tabular} & 7-22m & \begin{tabular}[c]{@{}c@{}}100-480bps\end{tabular} & \begin{tabular}[c]{@{}c@{}}Phone w/\\FM radio\end{tabular} \\ \midrule[0.5mm]
\begin{tabular}[c]{@{}c@{}}SideComm~\cite{SideComm}\\ (Cross-LoRa)\end{tabular} & Processors & 10-15m & 1kbps & SDR \\ \hline
\begin{tabular}[c]{@{}c@{}}EMLoRa~\cite{EMLoRa}\\ (Cross-LoRa)\end{tabular} & DRAM & 40-137m & \begin{tabular}[c]{@{}c@{}}1.25-14bps\end{tabular} & SDR \\ \hline
\begin{tabular}[c]{@{}c@{}}{\ourprotocol}\\ (Cross-LoRa)\end{tabular} & \begin{tabular}[c]{@{}c@{}}Video Cable\end{tabular} & \begin{tabular}[c]{@{}c@{}}40-132m\end{tabular} & \begin{tabular}[c]{@{}c@{}}180-21.6kbps\end{tabular} & \begin{tabular}[c]{@{}c@{}}\textbf{COTS LoRa}\\ \textbf{node/gateway}\\ \textbf{or SDR} \end{tabular}         \\ \bottomrule[0.5mm]
\end{tabular}
}
\label{Tab:RelatedWork}
\end{table}

\textbf{TEMPEST} (\underline{T}ransient \underline{E}lectro\underline{M}agnetic \underline{P}ulse \underline{E}manation \underline{ST}andard \\~\cite{enwiki:1073264081}) was established by the NSA and NATO in response to the concerns about the acceptable level of EMR from computers. Van Eck~\cite{Van1985} first demonstrated that attackers can recover the displayed content by analyzing the EMR of a TV. Such passive eavesdropping~\cite{EMeye,PhoneTEMPEST,TEMPESTSDR,sayakkara2018accuracy,gr-tempest,hayashi2014threat} is collectively referred to as TEMPEST attacks. Soft-TEMPEST~\cite{SoftTEMPEST} extended this concept by actively generating controlled EM emissions through software-layer operation. For example, by displaying carefully crafted black-white images on a monitor, the video cable can emit signals at specific frequencies.

\textbf{EM covert channels}~\cite{PhysicalCovertChannel,Bridgeware} are designed to modulate the EMR emitted from electronic devices to exfiltrate confidential information, without relying on conventional communication media (WiFi, Bluetooth, Internet, \etc). Previous works explored various manipulable leakage sources as summarized in Table~\ref{Tab:RelatedWork}.
GSMem~\cite{GSMem} manipulates the EM emission of the memory bus (DRAM) using specific memory-related CPU instructions, modulating secret data using Binary Amplitude Shift Keying (B-ASK) and sending it to a nearby (within 5.5m) spy phone with a rootkit implanted in its baseband firmware. BitJabber~\cite{Bitjabber} improves the data rates of the DRAM-related EM covert channel, achieving a throughput of 100-300 kbps within a 3m distance (its modulation methods include Binary Frequency Shift Keying and Multiple Frequency Shift Keying).
Noise-SDR~\cite{Noise-SDR} focuses on the customizability of DRAM's EMR using Radio-Frequency Pulse-Width Modulation (RF-PWM), and demonstrates receiving EM emission signals with multiple modulation modes on a USRP B210.
Air-Fi~\cite{Air-Fi} manipulates DRAM to emit binary data modulated with On-Off Keying (OOK modulation) on the 2.4 GHz band, requiring prior modifications for the WiFi adapter's driver and firmware.
`TEMPEST for Eliza'~\cite{TEMPESTForEliza} demonstrated that crafted screen images can generate intentional EM emission from VGA cables. While initially designed to transmit AM-modulated music via VGA cable's EMR, it provided foundational insights into image-based EMR signal modulation.
AirHopper~\cite{AirHopper} encodes data into Frequency-Modulation (FM) radio signals by manipulating VGA or HDMI cable's EMR via crafted screen images, using Audio Frequency-Shift Keying (A-FSK) or Dual-Tone Multiple-Frequency (DTMF) audio modulation to transmit data to nearby smartphones with FM receivers.
SideComm~\cite{SideComm} and EMLoRa~\cite{EMLoRa} are recent advances in cross-LoRa side-channel communication. EMLoRa is the first study to integrate the DRAM's EMR with CSS modulation, significantly extending the attack range, albeit at the cost of low transmission speeds.
SideComm focuses on using processors' EMR to provide additional wireless communication capabilities for low-power IoT devices. However, due to the limited EMR modulation capability (e.g., DRAM's EMR frequency cannot be modulated to the LoRa bands), the system implementations in all the above studies rely on specialized receivers such as SDRs.

\textbf{Cross-technology communication (CTC)} aims to achieve transparent transmission between incompatible wireless communication devices / protocols without the need for additional radio modules. Examples include WEBee~\cite{Webee} (WiFi to ZigBee), Bluebee~\cite{Bluebee} (BLE to ZigBee), BlueFi~\cite{BlueFi} (BLE to WiFi), L2X~\cite{L2X} and ZIMO~\cite{ZIMO}, which achieve CTC between different IoT connectivity technologies. Similarly, cross-LoRa studies such as LoRaBee~\cite{Lorabee} (LoRa to ZigBee), BLE2LoRa~\cite{BLE2LoRa} (BLE to LoRa), WiRa and Wi-Lo~\cite{WiRa,Wi-Lo} (WiFi to LoRa) facilitate interactions between LoRa and other wireless systems.

In contrast, \ourprotocol enables attackers to receive secret data through COTS LoRa nodes or gateways, outperforming most previous EM covert channels in both attack distance and data rates. To our knowledge, \ourprotocol is the first to integrate traditional CTC techniques with EM covert channels, achieving full compatibility of EMR with commercial wireless protocols.

\section{Overview}

\begin{figure}[t]
\centering
\includegraphics[width=0.95\linewidth]{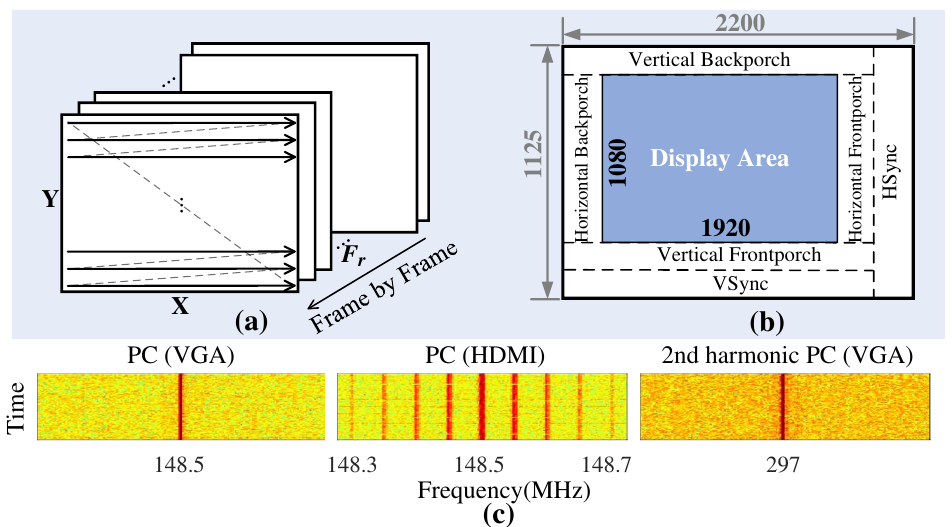}
\caption{(a) Monitor's line-by-line scanning manner. (b) The screen consists of a display area and some `hidden pixels'. (c) EMR of video cables at PC frequency and harmonics.}
\label{fig:LCD}
\end{figure}

\subsection{Background}
\label{sec:EMLeakageBackground}
\textbf{EM leakage of video cable}.
Monitors display images through a process of scanning and refreshing pixels. As depicted in Figure~\ref{fig:LCD} (a), when an image is transmitted to the monitor through a video cable, the monitor displays the pixels line-by-line from the top-left to the bottom-right corner and refreshes the entire screen at a frame rate $F_r$. Additionally, it includes a hidden display area at the edge of the screen, which is used for pixel synchronization and blanking~\cite{Hidden_pixels}, as shown in Figure~\ref{fig:LCD} (b).

Suppose each frame contains $Y$ scanlines, with each scanline consisting of $X$ pixels. The duration of one pixel $T_p$ is:
\begin{equation}
   T_p = \frac{1}{X \cdot Y \cdot F_r} 
\label{Equ:PixelTime}
\end{equation}

While the monitor is in operation, the video cable will emit EMR at the Pixel Clock (PC) frequency and its harmonics as shown in Figure~\ref{fig:LCD} (c) (these EMR spectra were captured while the monitor was displaying a black image). The value of PC can be calculated as:
\begin{equation}
    PC = \frac{1}{T_p}
\end{equation}

In this paper, we take the typical 1080X1920@60Hz display setting as a representative case for analysis. This setting corresponds to 1125 vertical lines and 2200 horizontal pixels per line, resulting in PC of 148.5 MHz~\cite{VESA}. In addition, because the HDMI employs Transition Minimize Differential Signaling (TMDS)~\cite{HDMIStd} for pixel encoding, HDMI's leakage spectrum contains more frequency components~\cite{TMDSLeakageAnalysis}.

\begin{figure}[t]
\centering
\includegraphics[width=0.95\linewidth]{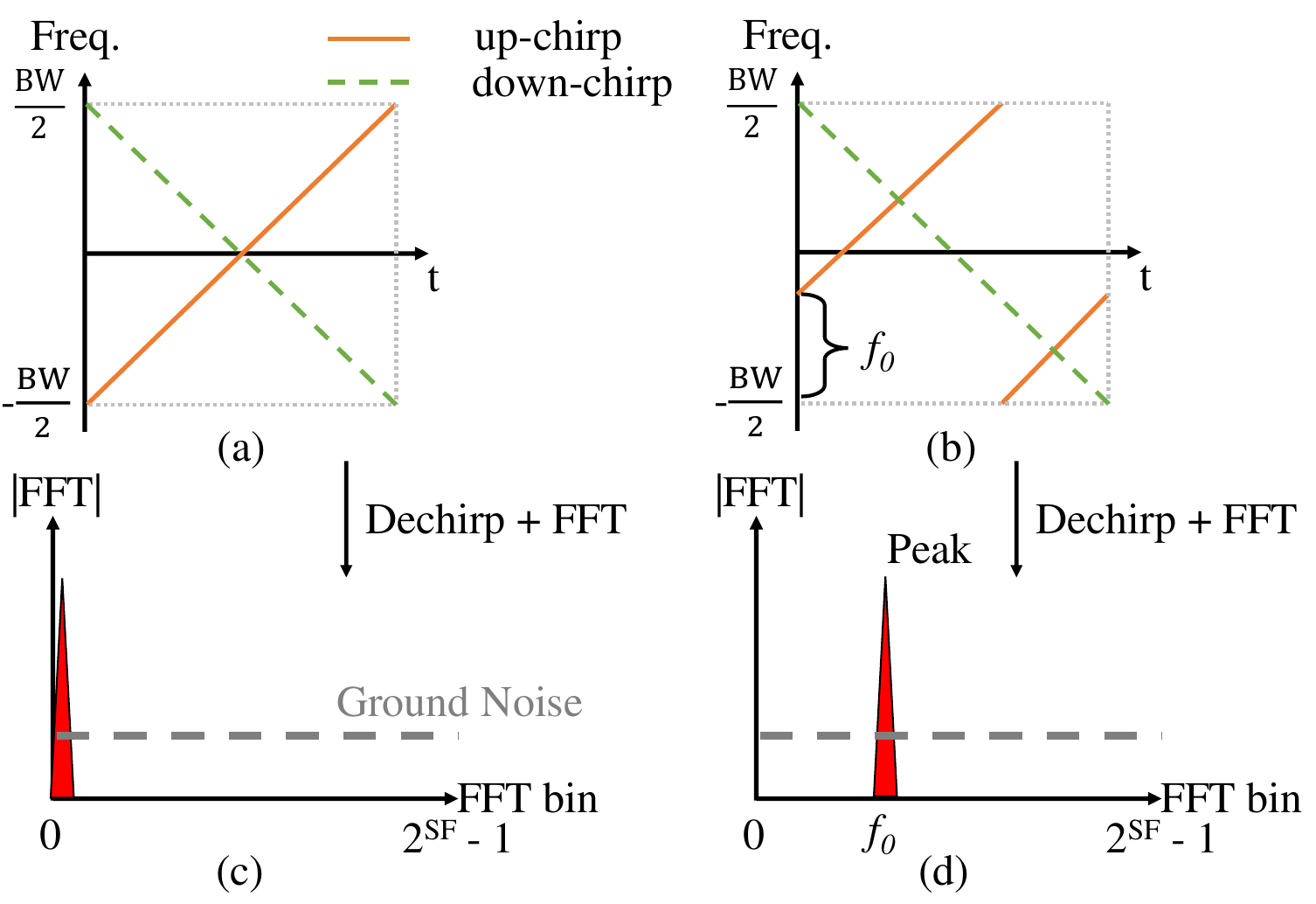}
\caption{(a) Basic up-chirp and basic down-chirp. (b) An up-chirp with $f_0$ shifting. (c) Demodulation for the basic up-chirp. (d) Demodulation for the up-chirp with $f_0$ shifting.}
\label{fig:CSSModulation}
\end{figure}

\textbf{LoRa physical layer}.
LoRa features long-range, low-power and high-sensitivity communication, which has been widely deployed all around the world to support various Internet of Things (IoT) applications~\cite{jouhari2023survey} such as environment monitoring, pet tracking, logistics and so on~\cite{yang2023aquahelper, xia2023xcopy}.

CSS modulation: LoRa's physical layer employs Chirp Spread Spectrum (CSS) modulation. A basic up-chirp (as shown in Figure~\ref{fig:CSSModulation} (a)) whose bandwidth spans from -$\frac{BW}{2}$ to $\frac{BW}{2}$ can be represented as:
\begin{equation}
    Upchirp(t) = e^{j2 \pi t( -\frac{BW}{2} + \frac{BW}{2T}t)}
\end{equation}
where $T$ denotes chirp duration. CSS modulates symbols by cyclically shifting the initial frequency $f_0$ as illustrated in Figure \ref{fig:CSSModulation} (b):
\begin{equation}
    y_e = Upchirp(t)e^{j2 \pi f_0 t}
\label{Equ:ShiftChirp}
\end{equation}
within the $BW$ range, up-chirps that commence at distinct initial frequencies are mapped to unique symbols. In the LoRa standard, $BW$ is partitioned into $2^{SF}$ distinct initial frequencies to encode $SF$-bits of data, where $SF$ refers to the \textit{spreading factor}. The most prevalent BW values for COTS LoRa are 125 kHz, 250 kHz, and 500 kHz, with $SF$ varying from 6 to 12.

For demodulation, a LoRa receiver accumulates energy through coherent de-spreading. The coherent signal down-chirp can be expressed as $Downchirp(t)=e^{j2 \pi t( \frac{BW}{2} - \frac{BW}{2T}t)}$. By utilizing the coherence of up-chirp and down-chirp, the result of multiplying the modulated up-chirp and the basic down-chirp is a single-frequency signal:
\begin{equation}
    Dechirp(t) = y_e \cdot Downchirp(t) = e^{j2 \pi f_0 t}
\end{equation}

Next, the receiver performs a Fast Fourier Transform (FFT) on $Dechirp(t)$, producing a power peak at frequency $f_0$. The index of this peak corresponds to the initial frequency of the coded symbol, as shown in Figure~\ref{fig:CSSModulation} (d).

In this paper, similar to the Signal-to-Noise Ratio (SNR), we define the dechirp-to-noise ratio (DNR) to quantify the signal quality of the demodulated chirp signal:
\begin{equation}
DNR = 20 \cdot \lg \frac{Peak}{Noise}
\label{Equ:DechirpGain}
\end{equation}
where $Peak$ is the peak value of the dechirp + FFT, and $Noise$ is the noise level. A higher DNR indicates superior signal quality.

\subsection{Attack Model}

\textbf{Victim}: We assume that the target is a computer storing confidential data of interest to the attacker, typically located in high-security environments such as isolated internal networks.
To protect against cyber-attacks, the victim's air-gapped computers have removed conventional communication modules (\eg, Wi-Fi, Bluetooth, and Ethernet). The victim's computer only retains essential components, including the host and the monitor with a VGA or HDMI cable. 

LoRa technology has been widely deployed in both indoor and outdoor IoT scenarios for data transmission~\cite{kolobe2020systematic}, environmental sensing~\cite{deng2020novel}, industry control~\cite{leonardi2019rt}, \etc. 
We assume the presence of LoRa nodes or gateways near the victim. 
This assumption is based on the fact that real-world standards for constructing air-gapped networks in various critical infrastructures, such as governments~\cite{barker2018recommendation,scarfone2008technical,pub1994security,force2017security}, the European Union~\cite{anna2020directive,ECSSecurity1,ECSSecurity2,GDPR}, businesses~\cite{GoogleStd,lai2009implementation}, industry~\cite{stouffer2011guide,knapp2024industrial,lendvay2016shadows}, medical~\cite{arendt2016medical}, and military sectors~\cite{chua2005cyberciege}, primarily focus on strictly disconnecting the external communication interfaces on air-gapped computers. These air-gapped systems have not yet delineated essential procedures for the elimination or segregation of commercial wireless devices within their vicinity (\eg, around 100m). 

\textbf{Attacker}: Consistent with previous EM covert channels~\cite{Bridgeware}, we assume that an attacker has implanted malware carrying \ourprotocol on the victim's computer through supply-chain attacks~\cite{Stuxnet,fireeye2020highly,SolarWinds,Ramsay} or social engineering tactics~\cite{Social1,Social2}. Once implanted, the malware scans the computer to locate confidential files and encrypts the secret data using the attacker's private key to ensure confidentiality. The attacker then covertly generates and plays an attack video when the computer is ensured to be unattended (such as in standby mode). The malware can obtain the resolution and refresh rate of the victim's monitor (\eg, 'xrandr' command in Linux) and generate the corresponding attack video.

For the receiving end, \ourprotocol considers two possible approaches:

\textbf{Approach 1: Receiving via operational COTS LoRa Gateways/Nodes}. The attackers can deploy their own COTS LoRa nodes or gateways near the air-gapped systems, or leverage third-party operational LoRa gateways deployed worldwide to receive EM packets. 

Subsequently, as shown in Figure~\ref{fig:AttackTarget}, the EM packets carrying sensitive data are transmitted to nearby LoRa gateways. The (third-party) LoRa receivers can then forward the LoRa packet to the attacker's application server in the cloud. A unique feature of \ourprotocol is its fine-grained control of EMR signals, which allows \ourprotocol to generate EM packets compatible with all LoRa packet configurations (all combinations of BWs and SFs). Thus, all deployed COTS LoRa devices could potentially receive and relay the covert packets with suitable parameter configurations to attackers worldwide. 

\textbf{Approach 2: Receiving with SDR on Flexible Spectrum}. 
Similarly to previous EM covert channels~\cite{EMLoRa}, attackers can also use low-cost SDRs (\eg, HackRF) to receive EM packets. \ourprotocol has a strong frequency-modulation capability for EMR (within 10 MHz to 1000 MHz). We demonstrate that attackers can select less-crowded frequency bands and customize LoRa-like (but different) EM packets to achieve even longer attack distances at higher data rates than LoRa-compatible EM packets. 

\textbf{Visual covertness}: To ensure visual covertness, we assume attackers can use DDCcontrol~\cite{DDCcontrol} to turn off the screen and keep the video cable continuously emitting EM packets. Specifically, attackers can first use \textit{ddccontrol -p} command to obtain the device code of the monitor, such as \textit{dev:/dev/i2c-3}. Subsequently, the attackers can modify the 0xe1 register address of the monitor to '1' to turn off the screen: \textit{ddccontrol -r 0xe1 dev:/dev/i2c-3 -w 1}. The 0xe1 address corresponds to the power supply status. When the value of 0xe1 is modified to 1, it prevents the monitor from refreshing the screen and shows a black-screen, while keeping the video cable transmitting the pixel stream and emitting EM packets. This capability enables attackers to launch covert transmissions during periods of inactivity, such as after office hours or when the system is left unattended.

\section{\ourprotocol}
\label{Sec:Attack}
This section presents the technical details of \ourprotocol. First, we discuss the EMR model of video cables and how to manipulate the EMR through customized attack images. Second, we elaborate how to construct EM packets compatible with the LoRa protocol, which can be directly received by operational COTS LoRa devices; and then, we present an enhanced version with low-cost SDRs.

\subsection{Video Cable EMR Transmitter}
\label{Sec:EMTransmitter}
We focus on the EMR inherently caused by the voltage fluctuations on video cables (VGA and HDMI). These fluctuations mainly stem from the operation of data buses responsible for transmitting pixel information and the clock buses that ensure synchronization. 

\begin{figure}[t]
\centering
\includegraphics[width=0.98\linewidth]{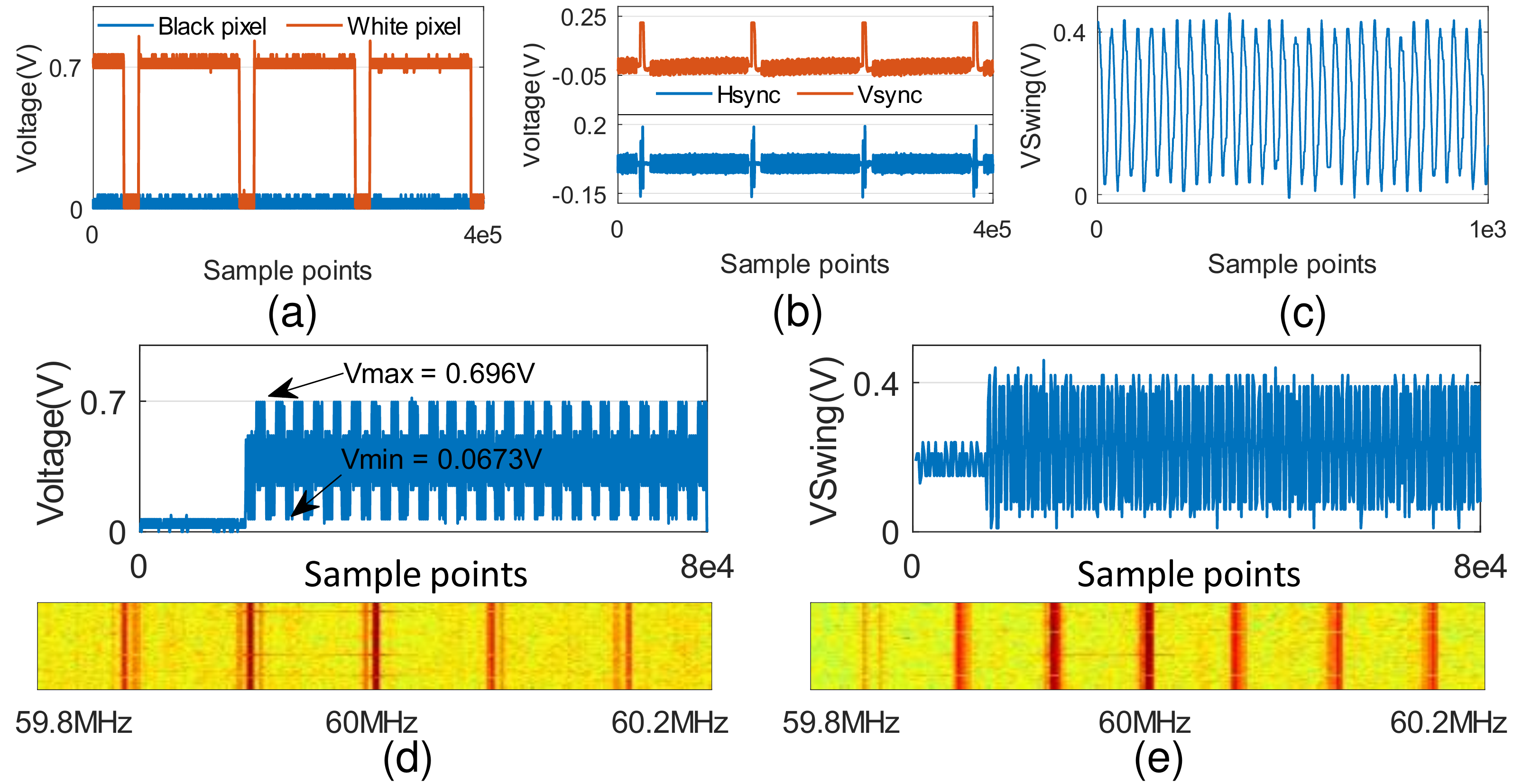}
\caption{(a). Voltage of VGA's blue bus when transmitting black/white pixels. (b). Voltage of VGA's Hsync and Ysync buses. (c). The voltage swing of HDMI's clock bus. (d) and (e). Voltage of VGA's blue bus and Voltage swing of HDMI's DATA1+- buses when displaying a 60 MHz attack image and their EMR spectra.}
\label{fig:Waveforms}
\end{figure}

VGA uses [R,G,B] data buses to transmit red, green and blue pixels in the form of analog signals, and HDMI has three pairs of differential data buses (DATA0+-, DATA1+-, and DATA2+-) for pixel transmission with digital signals. For clock synchronization, VGA's VSync and HSync buses are used for vertical and horizontal pixel synchronization, respectively, while HDMI employs a pair of independent clock+- differential buses.

To visualize the relationship between EMR and various bus activities, we use an oscilloscope (RIGOL Oscilloscope MSO5000) to measure the voltage waveforms of the data buses and clock buses of VGA and HDMI. It should be noted that when transmitting black [0,0,0] or white [1,1,1] pixels, VGA's R, G, and B buses exhibit identical voltage waveforms; The same applies to HDMI's three pair of differential data buses (DATA0+-, DATA1+-, and DATA2+-). Figure~\ref{fig:Waveforms} (a) illustrates the voltage waveforms on the Blue bus when transmitting 3 rows of black or white pixels, respectively. When scanning to the Porch and Sync areas, the voltage of the data bus is 0V (same as the voltage of the black pixel). Figure~\ref{fig:Waveforms} (b) is the voltage waveforms of VGA's Vsync and Hsync buses, and Figure~\ref{fig:Waveforms} (c) is the voltage waveform of the clock bus of HDMI. Since the bus speeds of the data buses and clock buses are both equal to the Pixel Clock (PC) frequency, the EMR frequencies of the VGA and HDMI cables are mainly concentrated at \textit{PC and its harmonics} as shown in Figure~\ref{fig:LCD} (c).

\textbf{Modeling:} We model the EMR according to their sources as follows:
\begin{equation}
    S_{sum} = S_{color} + S_{sync} + S_{other}
\label{Equ:Ssum}
\end{equation}
where $S_{color}$ is the aggregate EMR of all data buses that transmit pixel colors, $S_{sync}$ is the sum of the EMR of clock buses, and $S_{other}$ is the EMR collection of other sources (such as the GND bus). We further measure the voltage waveforms of other buses and confirm that the EMR signals come mainly from $S_{color}$ and $S_{sync}$, so we ignore $S_{other}$ in subsequent analysis.

Fine-grained control over the EMR source is essential to establish an EM covert channel. $S_{sync}$ comes from the clock bus that cannot be manipulated and constantly appears at the PC and its harmonics. In contrast, $S_{color}$, which comes from the data bus that transmits image information, is adjustable. Next, we focus on designing the pixel stream of attack images to modulate the EMR of $S_{color}$.

\begin{figure}
    \centering
    \includegraphics[width=0.99\linewidth]{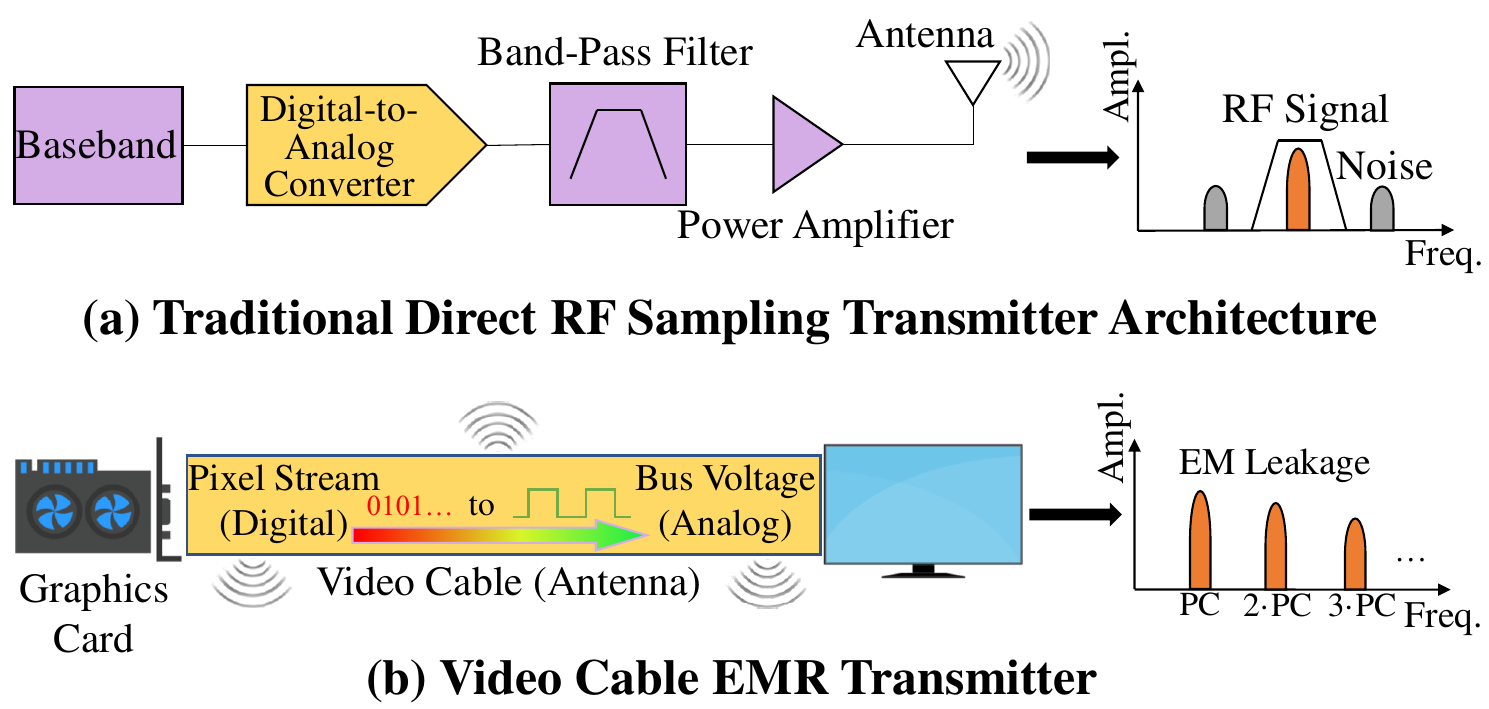}
    \caption{Contrast of direct RF sampling transmitter architecture and video cable EMR transmitter.} 
    \label{fig:RFTransmitter}
\end{figure}

\textbf{Video cable EMR transmitter:} From the radio-frequency (RF) transmitter perspective, the principle of EMR is similar to that of a 'direct RF sampling transmitter' (a typical RF hardware architecture that directly digitizes high-frequency RF signals~\cite{Direct-RF}). Specifically, during the transmission of image data from the graphics card to the monitor via the video cable, the digital pixel stream is converted to the analog voltage, and the color of pixel determines the voltage level on the data bus (\ie, the pixel stream corresponds to the baseband of the video cable). This conversion process from the pixel stream to the bus voltage largely fulfills the functionality of the Digital-to-Analog Converter (DAC) module in a direct RF sampling transmitter that converts the digital signal from the baseband to an analog signal, as shown in Figure~\ref{fig:RFTransmitter} (a). The video cable EMR transmitter operates at a sampling rate of PC, and the emitted EMR is directly determined by the frequency components of the voltages on the cable's buses, with the cable itself serving as the antenna. The difference is that the band-pass filter in the direct RF sampling transmitter can filter out noise outside the desired frequency band, and the power amplifier can boost the power of the RF signal. In contrast, the video cable EMR transmitter lacks such filtering, resulting in EMR at multiple harmonic frequencies of the PC, as illustrated in Figure~\ref{fig:RFTransmitter} (b).

To actively manipulate the frequency of EMR, the key idea is to emulate down-sampling of the pixel rate $S_{color}$ from PC to an expected frequency. Specifically, by carefully designing the pixel arrangement within the pixel stream to adjust the pixel frequency, the voltage frequency on the data bus can be indirectly modulated. When the frequency of bus voltage is the same as the expected EM leakage frequency, the video cable will emit EMR of the corresponding frequency. Algorithm~\ref{Alg:EMControl} outlines the EMR frequency modulation process in the attack image.

\begin{algorithm}[t]
\centering
\floatname{algorithm}{Algorithm}
\caption{Algorithm for designing attack image}
\label{Alg:EMControl}
\begin{algorithmic}[1]
	\State \textbf{Input:} Expected attack frequencies $\{f_1, f_2, ... , f_n\}$, pixel duration of each frequency $\{t_1, t_2, ..., t_n\}$, pixel clock $PC$
    \State Timer = 1;
    \State PixelStream = [];
        \For{$k=1$ to $n$}
            \For{$i=1$ to $t_k$}
                \State DownSampRate = $mod(f_k, PC) / PC$;
                \State Val = sin(2.0 $\cdot$ $\pi$ $\cdot$ DownSampRate $\cdot$ Timer);
                \If{Val $\textgreater$ 0}
                    \State PixelStream(Timer) = 1; // White pixel
                \Else
                    \State PixelStream(Timer) = 0; // Black pixel
                \EndIf
                \State Timer++;
            \EndFor
        \EndFor
        \State PixelStream = [PixelStream FramePadding];
        \State Image = reshape(PixelStream, ScreenH, ScreenW);
        \State Save2Image(Image[DisplayArea]);
\end{algorithmic}
\end{algorithm}

Algorithm~\ref{Alg:EMControl} is designed to generate an attack image that can manipulate the video cable sequentially to emit EMR at $f_1$ to $f_n$ frequencies, with each frequency's emission lasting for $t_k$ $\cdot$ $T_p$. Lines 6 to 13 are the core of Algorithm~\ref{Alg:EMControl}. Line 6 calculates the ratio for downsampling $S_{color}$ from the PC to $f_k$. The mod($f_k$, PC) function serves to downsample the EMR harmonics when the expected frequency $f_k$ exceeds PC. Lines 7 to 13 compute the pixel sequence required for the video cable to emit a single-frequency emission at $f_k$; here, white and black pixels are strategically utilized to alter the voltage waveform on the data buses. The benefit of using a black-white pattern is that it stimulates the cable to emit EM emission with maximum intensity. The sine function in Line 7 is an example of modulating the EMR into a sine wave (it can be modified to perform other modulation patterns). If the length of PixelStream is less than one frame, Line 16 fills the end with black pixels to make it a complete frame. Line 17 transforms the 1D PixelStream into a 2D attack image. For 1080x1920 resolution, the screen height $ScreenH$ and width $ScreenW$ equal 1125 pixels and 2220 pixels~\cite{VESA}. Line 18 selects the pixels of the Display Area (1080x1920) from the Image matrix (1125x2200) and saves as an attack image.

Figures~\ref{fig:Waveforms} (d) and (e) show examples of the voltage waveforms of the VGA and HDMI's data bus when the monitor displays a 60 MHz attack image. For the emission spectrum, since the waveform of the black-white pixel pattern is similar to a square wave, there are some frequency components near the expected frequency (the intensity of these components is always weaker than the expected frequency). These frequency components can be weakened by modifying Line 9 to Line 11 in Algorithm~\ref{Alg:EMControl} to the grayscale value corresponding to \textit{Val} (thus making the voltage waveform closer to a sine wave), but at the same time we find that the EMR intensity at the expected frequency $f_k$ will also be weakened. Therefore, in practice, we still use the black-white pattern, focusing on the EMR at $f_k$ and disregarding the lower intensity frequency components.

\begin{figure}[t]
\centering
\includegraphics[width=0.95\linewidth]{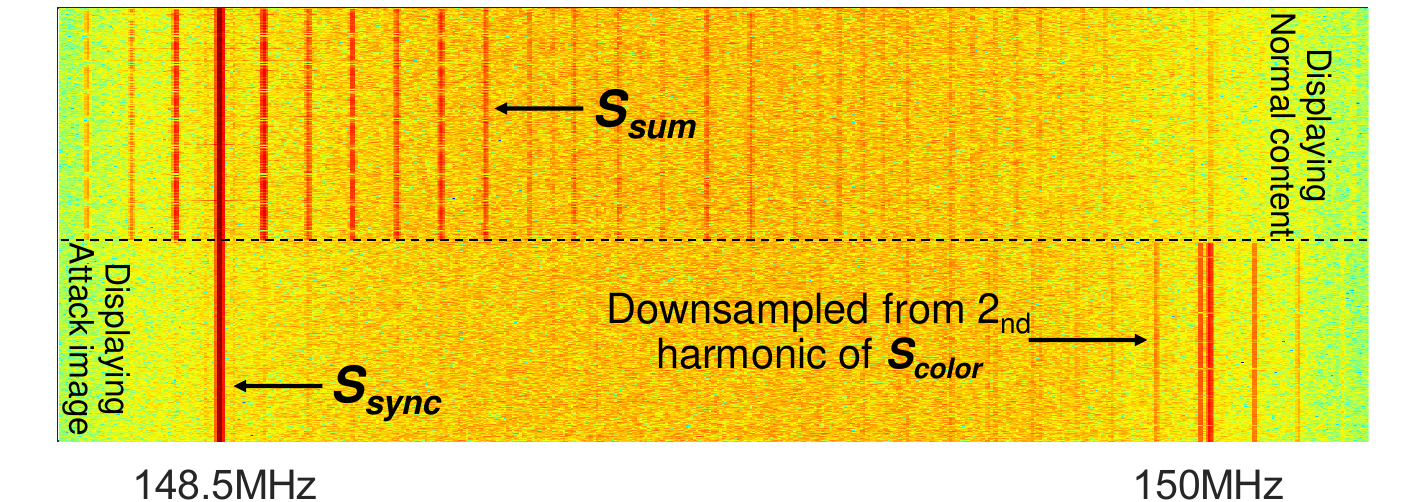}
\caption{Comparison of HDMI's EMR spectrum: normal emission and 150 MHz intentional emission.}
\label{fig:CompareEMR}
\end{figure}

\begin{figure}[t]
    \centering
    \includegraphics[width=0.99\linewidth]{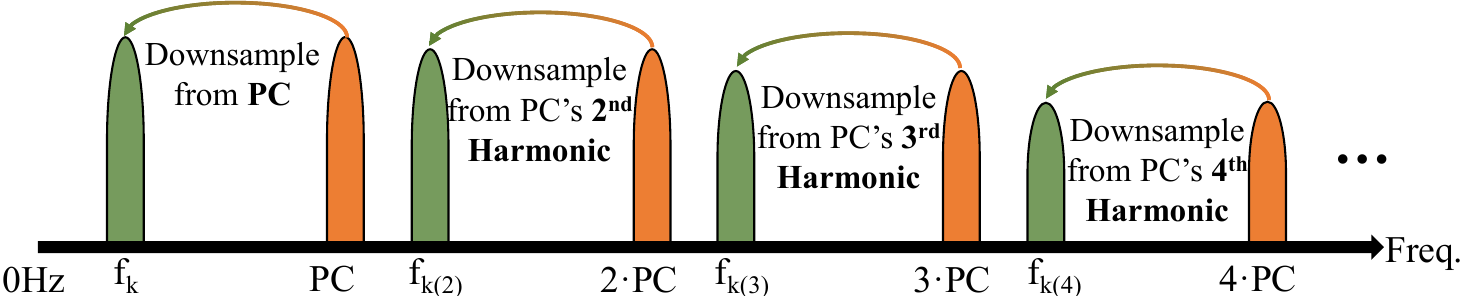}
    \caption{Emitting EMR at arbitrary frequency $f_k$ by downsampling PC or PC's $\lceil \frac{f_k}{PC} \rceil$-th harmonic.} 
\label{fig:DownSample}
\end{figure}

\begin{figure}[b]
\centering
\includegraphics[width=0.6\linewidth]{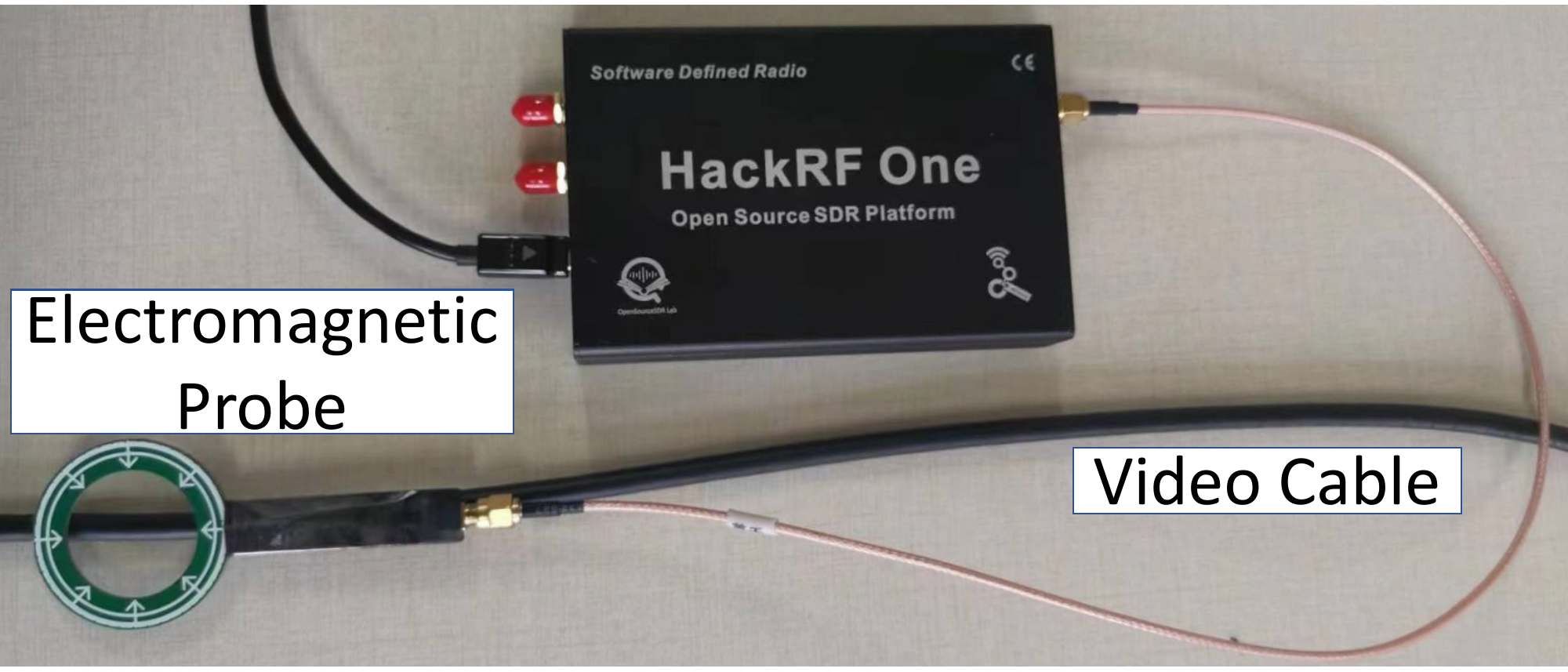}
\caption{We use HackRF with an EM probe to measure the EMR intensity of VGA and HDMI.}
\label{fig:HackRFandProbe}
\end{figure}

\begin{figure*}[t]
\centering
\includegraphics[width=0.98\linewidth]{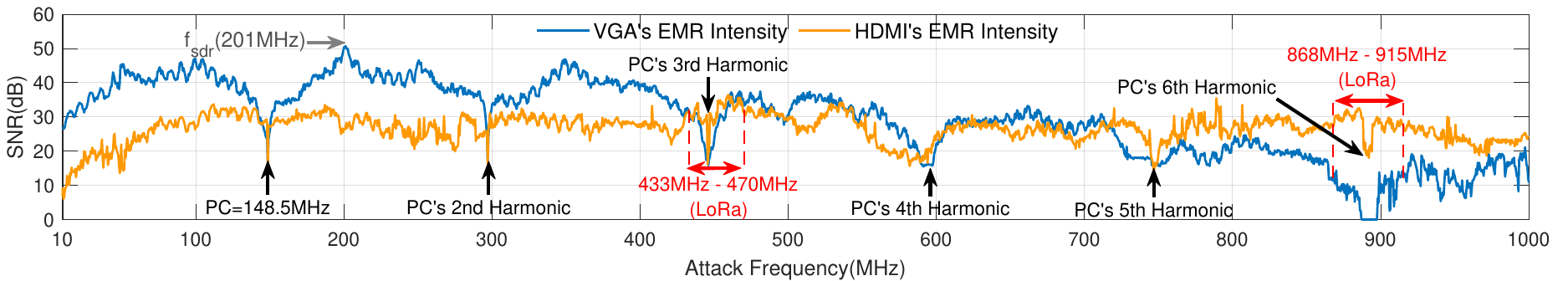}
\caption{Actual EMR intensity of VGA and HDMI with attack frequencies from 10 MHz to 1000 MHz.}
\label{fig:LeakSweep}
\end{figure*}

To illustrate how the attack image shifts the EMR's frequency, Figure~\ref{fig:CompareEMR} compares the HDMI leakage spectrum when no attack image is displayed and the 150 MHz attack image is displayed. The upper part is $S_{sum}$ near the PC frequency (148.5 MHz), and the monitor displayed a web page at this time. When the 150 MHz attack image is displayed, only the $S_{sync}$ leakage is left at the PC. The EMR that appears at 150 MHz is downsampled from the 2nd harmonic of $S_{color}$ (297 MHz) to 150 MHz (at the same time, the fundamental band of $S_{color}$ at 148.5 MHz is downsampled to 1.5 MHz). Building on this example, Figure~\ref{fig:DownSample} provides an abstract illustration of how EMR components at PC and its harmonics can be downsampled to an arbitrary target frequency $f_k$. When $f_k \in (0, \text{PC}]$, EMR is shifted from the PC frequency to $f_k$ by a specially crafted pixel stream that forms a strong EM emission component at $f_k$. (from the perspective of the RF transmitter, it is equivalent to reducing the baseband rate from PC to $f_k$). When $f_k \in (\text{PC}, +\infty)$, Algorithm~\ref{Alg:EMControl} targets the manipulation of $\lceil \frac{f_k}{\text{PC}} \rceil$-th harmonic of $S_{\text{color}}$. 

\textbf{Emission spectrum}. The main operating frequency bands of LoRa include 433 MHz - 470 MHz and 868 MHz - 915 MHz, which are within the 3rd - 7th harmonic range of EMR. To verify that the video cable EMR transmitter has the capability to be frequency compatible with LoRa protocol, we measured the SNR of EMR from VGA and HDMI cables using a HackRF~\cite{HackRF} with an electromagnetic probe as shown in Figure~\ref{fig:HackRFandProbe}, and the probe was tightly attached to the video cable. In an electromagnetic darkroom, we create and display attack images on the monitor, configured at 1080x1920@60Hz, with attack frequencies sweeping from 10 MHz to 1000 MHz. Each attack image is designed to emit one single attack frequency, with a frequency interval of 0.5 MHz between adjacent attack images.

Figure~\ref{fig:LeakSweep} illustrates VGA and HDMI's EMR intensity at attack frequencies ranging from 10 MHz to 1000 MHz. It is evident that VGA demonstrates significantly higher leakage intensity than HDMI within the frequency band below the PC's 3rd harmonic (10 MHz - 445.5 MHz). In contrast, above the PC's 5th harmonic (742.5 MHz - 1000 MHz), HDMI shows a slightly higher intensity than VGA.

The disparity in EMR intensity between VGA and HDMI is due to their voltage ranges and pixel encoding methods. In the low-frequency band, the EMR intensity is mainly determined by the magnitude of cable voltage fluctuation. VGA operates within a voltage range of 0V to 0.7V, whereas HDMI has a voltage swing of 0.4V, hence VGA's EMR intensity is higher than that of HDMI below PC Hz. However, the black-white pixel stream drives the voltage waveform of VGA to approximate an ideal square wave, which more effectively concentrates the emission energy in the lower frequency band. Due to TMDS encoding, HDMI disperses part of its energy across frequencies other than the target attack frequency, resulting in a wider emission spectrum~\cite{paul2022introduction} and a more evenly distributed EMR intensity across the 10-1000 MHz range compared to VGA.
In addition, we observed deviations between the actual bus voltage and the prescribed manufacturing standards. For example, when the monitor displays a 60 MHz attack image, the actual valley and peak voltages of the VGA are 0.0673V and 0.696V as illustrated in Figure~\ref{fig:Waveforms} (d), which are not strictly equal to the specified 0V and 0.7V. 
We speculate that the alternating black-and-white pixel pattern leads to voltage overshoot and undershoot~\cite{lin2010overshoot}, which in turn results in fluctuations in emission intensity across various attack frequencies.

Although VGA and HDMI exhibit sufficient emission intensity to achieve frequency compatibility with LoRa bands (433-470 MHz and 868-915 MHz), the SNR significantly degrades near the PC frequency and its harmonics. This decline is mainly due to interference from the EMR of $S_{sync}$. Therefore, to maintain covert channel quality, attackers should avoid selecting frequencies close to the 3rd and 6th harmonics of the PC.

\subsection{CTCC to LoRa}
\label{Sec:CTCC}
In this subsection, we present the methodology for constructing LoRa-compatible EM packets. Then, we explore strategies for performance enhancement that attackers could exploit with low-cost SDRs.

The core idea of modulating EMR into LoRa waveforms is to continuously shift the emission frequency in accordance with LoRa receiver's settings (SF and BW) at a selected LoRa frequency. This controlled frequency sweeping emulates the chirp signals used in LoRa modulation and enables the construction of LoRa-compatible EM packets.

\begin{figure}
    \centering
    \includegraphics[width=0.95\linewidth]{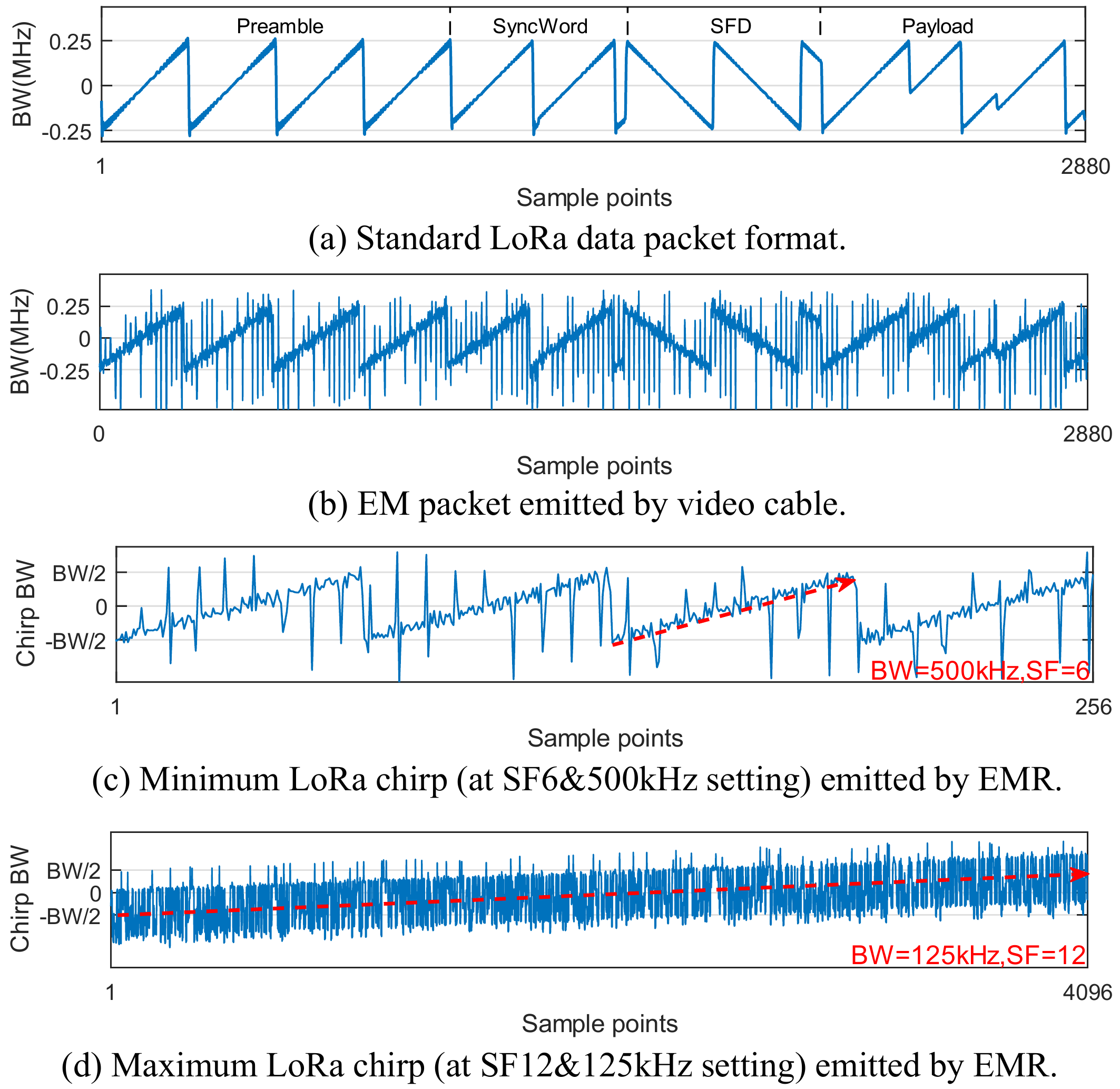}
    \caption{COTS LoRa signal waveform and EM chirps.} 
    \label{fig:PacketWave}
\end{figure}

Figure~\ref{fig:PacketWave} (a) illustrates a standard LoRa packet configured with SF=8\&BW=500kHz. The Preamble part comprises multiple consecutive basic up-chirps and helps the receiver detect the incoming signal and synchronize its timing; the SyncWord, made up of two up-chirps, serves to distinguish different LoRa networks; the Start Frame Delimiter (SFD) contains 2.25 basic down-chirps, which mark the beginning of the Payload part; finally, the Payload carries actual encoded data. To ensure that EM packets can be recognized and decoded by the COTS LoRa devices, the chirp duration and frequency band need to comply with LoRa standards.

\textbf{Mapping chirp duration to pixel number}. A LoRa chirp consists of $2^{SF}$ samples at a sampling rate of BW. Thus, the chirp duration can be calculated as follows:
\begin{equation}
T_{chirp} = \frac{2^{SF}}{BW}
\label{Equ:ChirpTime}
\end{equation}

Given that the sampling rate of the video cable EMR transmitter is equal to PC (Hz), generating a LoRa-style EM chirp requires $N_{pixel}$ consecutive pixels:
\begin{equation}
N_{pixel} = T_{chirp} \cdot PC
\label{Equ:NChirp}
\end{equation}

\textbf{EM chirp modulation}.
We modified the input and modulation parts of Algorithm~\ref{Alg:EMControl} to emit an EM chirp whose frequency varies linearly over the [$f_c$ - $\frac{BW}{2}$, $f_c$ + $\frac{BW}{2}$] range.

\begin{algorithm}[h]
\floatname{algorithm}{Algorithm}
\caption{For emitting a modulated up-chirp EMR}
\label{Alg:EMChirp}
\begin{algorithmic}[1]
	\State \textbf{Input:} $f_{low}$=$f_c$ - $\frac{BW}{2}$, $f_{high}$=$f_c$ + $\frac{BW}{2}$, pixel duration $N_{pixel}$, spreading factor $SF$,
 chirp symbol offset $K$, $PC$
\State ...
    \State $F1$ = $mod(f_{low}, PC) / PC$;
    \State $F2$ = $mod(f_{high}, PC) / PC$;
    \State $F_{step}$ = $(F2 - F1) / N_{pixel}$;
    \State Val = $\sin(2.0 \cdot \pi \cdot (F1 + F_{\text{step}} \cdot \text{mod}(N_{\text{pixel}} \cdot \frac{K}{2^{SF}} + \text{Timer} - 1, N_{\text{pixel}})) \cdot \text{Timer}$)
\State ...
\end{algorithmic}
\end{algorithm}

In Algorithm~\ref{Alg:EMChirp}, $f_{low}$ and $f_{high}$ define the frequency boundaries; $N_{pixel}$ is the pixel length required for the video cable to emit an EM chirp; $SF$ and $BW$ are consistent with the LoRa receiver settings; $K$ is the symbol offset corresponding to the chirp to be transmitted. Line 3 to line 5 calculate the downsampling ratios $F1$ and $F2$ to emit the EMR at $f_{low}$ and $f_{high}$ frequencies.
Line 6 modulates an EM chirp with an initial frequency of $f_{low}$ + $\frac{K}{2^{SF}} \cdot BW$ and $K$ is an integer between [0, $2^{SF}$-1]. For the SFD part, Algorithm~\ref{Alg:EMChirp} reverses $f_{low}$ and $f_{high}$ to emit the down-chirp.

To demonstrate the effectiveness of this approach, Figure~\ref{fig:PacketWave} (b) shows a high-fidelity EM packet emitted by a VGA cable, which closely mirrors the chirp duration and frequency of the standard LoRa packet in Figure~\ref{fig:PacketWave} (a). 
The noise within the EM packet is attributed to signal intervals when the monitor scans to the end of each scanline.
Due to LoRa's high sensitivity and strong noise resilience, COTS LoRa receivers can still reliably identify and decode these LoRa-compatible EM packets.

Since LoRa's configuration encompasses a rich combination of BW and SF, each combination requires the video cable to emit EM chirp signals based on different EMR shifting speeds and transmission durations. Specifically, the combination of SF6\&500 kHz requires the shortest chirp duration, while SF12\&125 kHz corresponds to the longest chirp duration, which are the upper and lower limits of all possible combinations. Fortunately, the video cable EMR transmitter's high sampling rate (148.5 MHz) provides sufficient resolution and timing precision to handle all such cases. Figures~\ref{fig:PacketWave} (c) and (d) demonstrate these two extremes via HDMI cable.

\begin{figure}[t]
    \begin{minipage}{0.95\linewidth}
        \subfigure[Frame interval destroys part of chirps of the EM packet.]{
                \centering
        \includegraphics[width=0.42\linewidth]{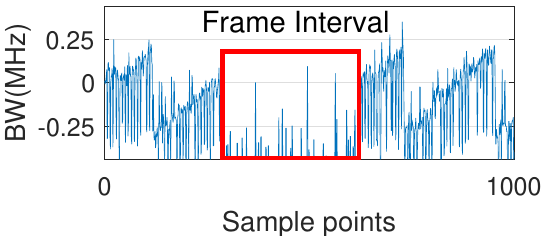}
        \label{fig:FrameInterval}
        }
        \subfigure[COTS LoRa devices may decode incorrectly due to frame interval.]{
                \centering
        \includegraphics[width=0.48\linewidth]{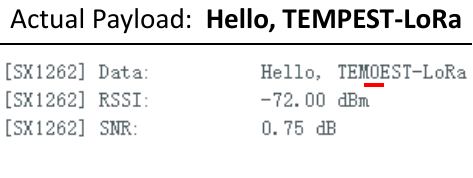}
        \label{fig:BrokenPacket}
        }
        \caption{Frame interval and broken EM packet.}
    \end{minipage}
\end{figure}

\begin{figure}
    \centering
    \includegraphics[width=0.95\linewidth]{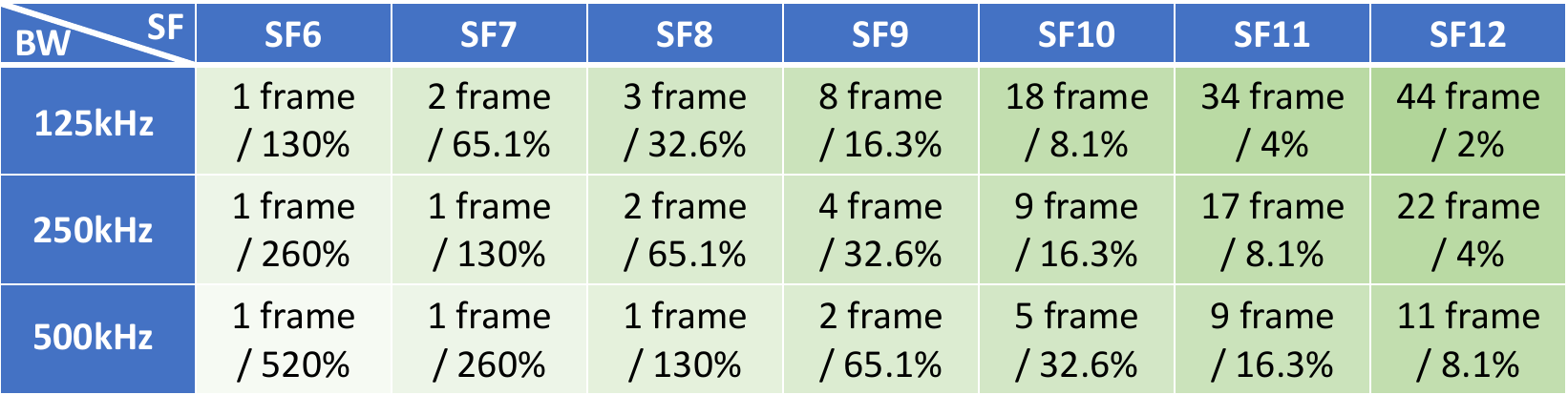}
    \caption{Minimum number of attack frames required assuming the payload size are 24 bits.} 
    \label{fig:FrameNumber}
\end{figure}

\textbf{Impact of frame interval}: When the duration of a single EM packet exceeds the time span of one frame (approximately 16.7ms), \ourprotocol combines multiple attack frames into an attack video to emit a longer EM packet.However, the unavoidable frame interval (the time interval between the end of one frame and the start of the next) introduces an emission gap of approximately 0.673ms, which may disrupt the continuity of the EM signal and cause the resulting packet to be decoded incorrectly. Figure~\ref{fig:FrameInterval} illustrates a segment of an EM packet affected by such a frame interval. In this example, the chirp is configured with SF6\&BW500 kHz. The intended LoRa-compatible EM packet carries the Payload of \textit{Hello, TEMPEST-LoRa}. However, due to the emission gap caused by the frame interval, the decoded output is \textit{Hello, TEMOEST-LoRa} as shown in Figure~\ref{fig:BrokenPacket}.

The settings of SF and BW, the number of frames, and the payload length jointly determine the effect of the frame interval on the EM packets. To provide an understanding of this trade-off, Figure~\ref{fig:FrameNumber} illustrates the number of frames required by \ourprotocol to construct an attack video under various combinations of SF (6 to 12) and BW (125, 250, and 500 kHz), assuming a fixed preamble of 4 up-chirps and a payload length of 24 bits (raw data length). Since the duration of the frame interval is fixed, Figure~\ref{fig:FrameNumber} also shows the ratio between the frame interval duration and the duration of the corresponding chirp. Increasing the chirp duration reduces the relative impact of frame intervals may corrupt EM packets, but at the cost of requiring more frames in the attack video. Note that since the LoRa physical layer uses error correction mechanisms such as the Hamming code, even EM packets containing frame intervals may be correctly decoded. We evaluated the packet reception rate under different parameter configurations in Section~\ref{Sec:PRR}.

\subsection{Low-cost SDR-based Enhancement}
The above section showed that attackers could generate LoRa-compatible EM packets and receive them with operational LoRa devices deployed worldwide. To exploit LoRa nodes/gateways, attackers are limited to the parameter configurations (\eg, SF, BW, central frequency, payload size, \etc) of the LoRa standard. 
In the following, we present an SDR enhanced \ourprotocol, which could enable attackers the increased flexibility in generating EM packets not limited by the commercial LoRa standard.

\textbf{Flexible frequency selection}. Given the flexible frequency selection capabilities of the video cable EMR transmitter and SDR, attackers will be capable of selecting and generating an arbitrary attack frequency within 1000 MHz to evade potential detection mechanism or select frequencies with higher intensity to enhance attack distance. We selected $f_{sdr}$ = 201 MHz as the representative frequency of this SDR-based version from Figure~\ref{fig:LeakSweep}. $f_{sdr}$ is away from the crowded ISM band, which can reduce interference from other coexisting wireless protocols (e.g. RFID~\cite{RFID433MHz}, ZigBee~\cite{ZigBee}, NB-IOT~\cite{NB-IoT-Protocol} and GSM~\cite{GSMProtocol}), and the risk of being detected by other COTS wireless receivers.

\begin{figure}
\centering
\includegraphics[width=0.9\linewidth]{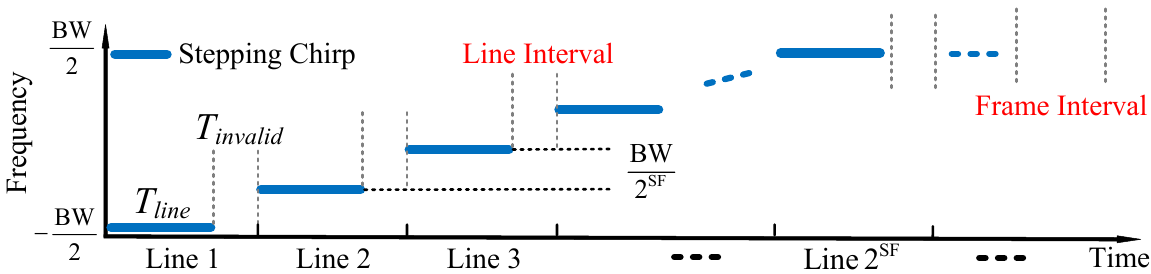}
\caption{Stepping chirp for align signal intervals. Line interval and frame interval are aligned within each step and each frame respectively.}
\label{fig:SteppingChirp}
\end{figure} 

\textbf{Align the signal interval}: To overcome the damage to the continuity of EM packets caused by the signal intervals, we customized a stepping chirp to align the line interval and frame interval as shown in Figure~\ref{fig:SteppingChirp}. Within one frame image, each scanline emits one single-frequency EMR, and consecutive $2^{SF}$ scanlines together constitute a stepping chirp EMR. In this way, the line interval is aligned in each step, and the frame interval is aligned in each frame, thus ensuring the signal continuity of the EM packets. The frequency separation between two adjacent steps is $\frac{BW}{2^{SF}}$. Each line step $T_{total}$ comprises the duration $T_{line}$ of one scanline within display area and the line interval $T_{invalid}$, which can be expressed as:
\begin{equation}
\begin{aligned}
&T_{total} = T_{line} + T_{invalid}, \\
&T_{line} = T_p \cdot X_{pixel}, & T_{invalid} = T_p \cdot X_{invalid}
\label{Equ:Tline}
\end{aligned}
\end{equation}
here, $X_{pixel}$ represents the horizontal pixel count (1920 pixels) within the display area, while $X_{invalid}$ denotes the aggregate pixel count of Porch and Sync regions with in a single scanline (2200 - 1920 pixels). Thus, one stepping up-chirp $UC_{step}(t)$ in Figure~\ref{fig:SteppingChirp} can be expressed as:

\begin{small}
\begin{equation}
UC_{\text{step}}(t) = \sum_{i=0}^{2^{SF}-1} \sin\left(2\pi \left(f_c - \frac{BW}{2} + i \cdot \frac{BW}{2^{SF}}\right) t\right) \cdot 
\text{rect}\left(\frac{t - i \cdot T_{\text{total}}}{T_{\text{line}}}\right)
\end{equation}
\end{small}

\noindent where the rectangular window function is defined as
\[
\text{rect}(\tau) =
\begin{cases}
1, & -\frac{1}{2} \leq \tau \leq \frac{1}{2} \\
0, & \text{otherwise}
\end{cases}
\]

where $f_c$ is the chirp's center frequency, and $rect(\tau)$ is used to describe the transmission of the signal within a specific period of time. Specifically, for each step ($T_{total}$), the rect($\tau$) function acts as a switch only when $t$ is 'on' (value 1) in the period of $T_{line}$ and 'off' (value 0) in $T_{invalid}$. This ensures that the step frequency of each EMR is emitted only within the specified time slot, resulting in a stepping chirp signal. In the stepping dechirp calculation, the receiving end utilizes the basic stepping down-chirp $DC_{step}(t)$ to demodulate the received stepping up-chirp:
\begin{small}
\begin{equation}
DC_{step}(t) = \sum_{i=0}^{2^{SF}-1} \sin\left(2\pi \left(f_c + \frac{BW}{2} - i \cdot \frac{BW}{2^{SF}}\right) t\right) \cdot \text{rect}\left(\frac{t - i \cdot T_{\text{total}}}{T_{\text{line}}}\right)
\label{Equ:StairDownChirp}
\end{equation}
\end{small}

In this SDR-based version, we set the SF to 10, and the residual 56 scanlines (1080-$2^{10}$) and the frame interval serve as a guard interval. This stepping chirp's duration is approximately equal to the parameter configuration of SF = 12 and BW = 250 kHz in the standard LoRa protocol.

\begin{figure}[t]
\centering
\includegraphics[width=0.99\linewidth]{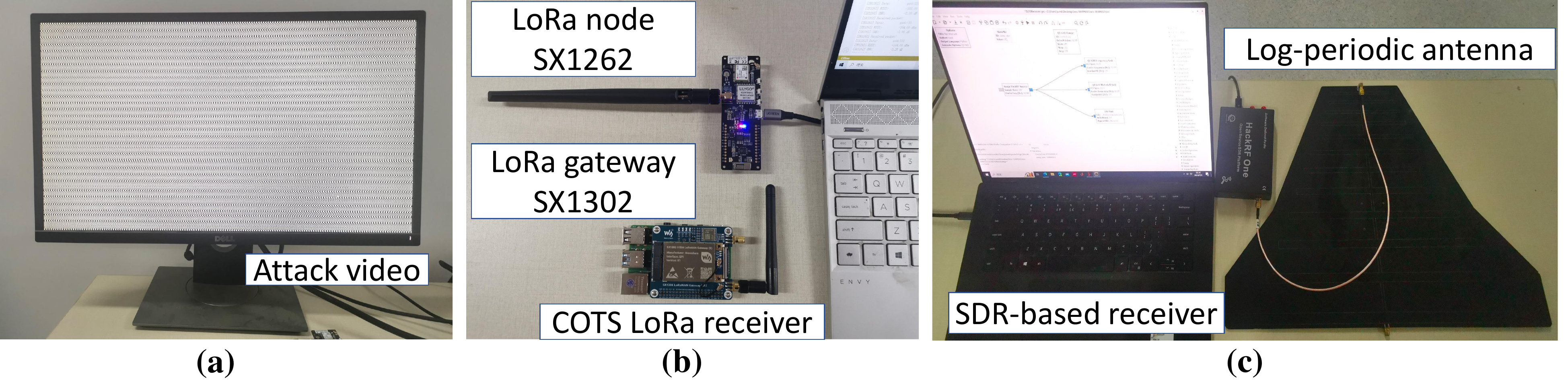}
\caption{(a) The monitor displaying the attack image/video. (b) COTS LoRa node (SX1262) and gateway (SX1302) with LoRa antennas. (c) SDR-based receiver.}
\label{fig:AttackSetup}
\end{figure}

\section{Experimental Evaluation}
\textbf{Experiment setup}: Except for the cross-device evaluation in Section~\ref{Sec:Cross-device}, we use a DELL P2317H monitor connected via a 1.5m VGA or HDMI1.4 cable (manufactured by UGREEN) to play the attack videos. The monitor is configured with a resolution of 1080x1920 and a refresh rate of 60 Hz, as illustrated in Figure~\ref{fig:AttackSetup} (a).

For COTS LoRa receivers, we use SX1262 LoRa nodes (for evaluation at 433 MHz) from LILYGO~\cite{Lilygo} and SX1302 LoRa gateways (for evaluation at 915 MHz) from WaveShare~\cite{WaveShare}, equipped with standard circular antennas (2.2 dBi gain) as shown in Figure~\ref{fig:AttackSetup} (b). Upon detecting EM packets, these devices return data (Payload) and Received Signal Strength Indicator (RSSI). For  SDR-based \ourprotocol, the chirp bandwidth is fixed at 500 kHz, and we employ a HackRF One~\cite{HackRF} paired with a log-periodic antenna (6 dBi gain) to capture the physical layer samples for processing, as shown in Figure~\ref{fig:AttackSetup} (c). 

\subsection{Cross-device feasibility}
\label{Sec:Cross-device}
To evaluate the cross-device feasibility of \ourprotocol, we use various commercially available display devices and video cables to emit EM packets at 433 MHz and 915 MHz. Then we record the RSSI values returned by the COTS LoRa devices.

\begin{table}[h]
\caption{Testing of \ourprotocol on different cable specifications.}
\resizebox{0.98\columnwidth}{!}{
\begin{tabular}{c|c|c|c|c}
\hline
\begin{tabular}[c]{@{}c@{}}\textbf{Cable}\\ \textbf{Manuf.}\end{tabular} & \textbf{Type} & \textbf{Length}  & \begin{tabular}[c]{@{}c@{}}\textbf{Returned RSSI} \\ \textbf{at 433MHz}\end{tabular} & \begin{tabular}[c]{@{}c@{}}\textbf{Returned RSSI}\\ \textbf{at 915MHz}\end{tabular} \\ \hline
SAMZHE & \begin{tabular}[c]{@{}c@{}}VGA/\\ HDMI1.4\end{tabular} & 1.5m & \begin{tabular}[c]{@{}c@{}}-76 dBm/\\ -79 dBm\end{tabular} & \begin{tabular}[c]{@{}c@{}}-97 dBm/\\ -74 dBm\end{tabular} \\ \hline
PHILIPS & \begin{tabular}[c]{@{}c@{}}VGA/\\ HDMI1.4\end{tabular} & 1.5m & \begin{tabular}[c]{@{}c@{}}-76 dBm/\\ -77 dBm\end{tabular} & \begin{tabular}[c]{@{}c@{}}-96 dBm/\\ -74 dBm\end{tabular} \\ \hline
CHOSEAL & \begin{tabular}[c]{@{}c@{}}VGA/\\ HDMI1.4\end{tabular} & 1.5m & \begin{tabular}[c]{@{}c@{}}-74 dBm/\\ -76 dBm\end{tabular} & \begin{tabular}[c]{@{}c@{}}-93 dBm/\\ -76 dBm\end{tabular} \\ \hline
HP & \begin{tabular}[c]{@{}c@{}}VGA/\\ HDMI1.4\end{tabular} & 1.5m & \begin{tabular}[c]{@{}c@{}}-73 dBm/\\ -75 dBm\end{tabular} & \begin{tabular}[c]{@{}c@{}}-94 dBm/\\ -76 dBm\end{tabular} \\ \hline
\multirow{4}{*}{UGREEN} & \begin{tabular}[c]{@{}c@{}}VGA/\\ HDMI1.4\end{tabular} & 0.5m & \begin{tabular}[c]{@{}c@{}}-78 dBm/\\ -78 dBm\end{tabular} & \begin{tabular}[c]{@{}c@{}}-95 dBm/\\ -75 dBm\end{tabular} \\ \cline{2-5} 
 & \begin{tabular}[c]{@{}c@{}}VGA/\\ HDMI1.4/\\ HDMI2.0\end{tabular} & 1.5m & \begin{tabular}[c]{@{}c@{}}-78 dBm/\\ -77 dBm/\\ -77 dBm\end{tabular} & \begin{tabular}[c]{@{}c@{}}-95 dBm/\\ -75 dBm/\\ -76 dBm\end{tabular} \\ \cline{2-5} 
 & \begin{tabular}[c]{@{}c@{}}VGA/\\ HDMI1.4\end{tabular} & 5m & \begin{tabular}[c]{@{}c@{}}-76 dBm/\\ -75 dBm\end{tabular} & \begin{tabular}[c]{@{}c@{}}-94 dBm/\\ -75 dBm\end{tabular} \\ \cline{2-5} 
 & \begin{tabular}[c]{@{}c@{}}VGA/\\ HDMI1.4\end{tabular} & 10m & \begin{tabular}[c]{@{}c@{}}-73 dBm/\\ -73 dBm\end{tabular} & \begin{tabular}[c]{@{}c@{}}-92 dBm/\\ -73 dBm\end{tabular} \\ \hline
\multicolumn{5}{c}{This test used Dell P2317H monitor.}\\ \hline
\end{tabular}
}
\label{Tab:Cross-Cable}
\end{table}

We first tested video cables from different manufacturers, with cable types and lengths as shown in Table~\ref{Tab:Cross-Cable}. All cables have shielded metal layers (\ie, aluminum foil wrapped around each bus) to minimize EM leakage. The results show that all video cables tested were able to be manipulated by \ourprotocol to emit EM packets. The reason for \ourprotocol's cross-device compatibility is that the electrical characteristics of commercially available display devices and video cables all follow the manufacturing standards defined by the Video Electronics Standards Association (VESA)~\cite{VESA}, so their EM emission characteristics are also similar.

In general, there was no significant difference in the EMR intensity of the video cable of different manufacturers, and the corresponding RSSI values were similar. \ourprotocol is also compatible with VGA and two mainstream versions of HDMI (1.4 and 2.0). In the measurement of cable length, we tested VGA and HDMI1.4 from UGREEN with lengths of 0.5m, 1.5m, 5m, and 10m. We noticed that the RSSI value increased slightly with increasing cable length. For example, the RSSI corresponding to HDMI1.4 from 0.5m to 10m increased from -78 dBm to -73 dBm. We speculate that the actual power of the longer cable will increase slightly, resulting in an increase in EMR intensity.

\begin{table}[h]
\caption{Testing of \ourprotocol on different display devices.}
\resizebox{0.98\columnwidth}{!}{
\begin{tabular}{c|c|c|c|c}
\hline
\begin{tabular}[c]{@{}c@{}}\textbf{Device}\\ \textbf{Type}\end{tabular} & \begin{tabular}[c]{@{}c@{}}\textbf{Device} \\ \textbf{Manuf.}\end{tabular} & \textbf{Model} & \begin{tabular}[c]{@{}c@{}}\textbf{Returned RSSI}\\ \textbf{at 433MHz}\\ \textbf{(VGA/HDMI)}\end{tabular} & \begin{tabular}[c]{@{}c@{}}\textbf{Returned RSSI}\\ \textbf{at 915MHz}\\ \textbf{(VGA/HDMI)}\end{tabular} \\ \hline
\multirow{5}{*}{Monitor} & Dell & P2317H & \begin{tabular}[c]{@{}c@{}}-78 dBm/\\ -77 dBm\end{tabular} & \begin{tabular}[c]{@{}c@{}}-95 dBm/\\ -75 dBm\end{tabular} \\ \cline{2-5} 
 & Dell & P2225H & \begin{tabular}[c]{@{}c@{}}-80 dBm/\\ -77 dBm\end{tabular} & \begin{tabular}[c]{@{}c@{}}-96 dBm/\\ -76 dBm\end{tabular} \\ \cline{2-5} 
 & PHILIPS & 275M7C & \begin{tabular}[c]{@{}c@{}}-77 dBm/\\ -75 dBm\end{tabular} & \begin{tabular}[c]{@{}c@{}}-93 dBm/\\ -75 dBm\end{tabular} \\ \cline{2-5} 
 & HP & P24V G5 & \begin{tabular}[c]{@{}c@{}}-78 dBm/\\ -78 dBm\end{tabular} & \begin{tabular}[c]{@{}c@{}}-95 dBm/\\ -76 dBm\end{tabular} \\ \cline{2-5} 
 & Xiaomi & RMMNT27NF & \begin{tabular}[c]{@{}c@{}}-78 dBm/\\ -77 dBm\end{tabular} & \begin{tabular}[c]{@{}c@{}}-97 dBm/\\ -77 dBm\end{tabular} \\ \hline
\multirow{2}{*}{Projector} & SONY & VPL-U300WZ & \begin{tabular}[c]{@{}c@{}}-75 dBm/\\ -76 dBm\end{tabular} & \begin{tabular}[c]{@{}c@{}}-93 dBm/\\ -74 dBm\end{tabular} \\ \cline{2-5} 
 & EPSON & CB-X05E & \begin{tabular}[c]{@{}c@{}}-80 dBm/\\ -77 dBm\end{tabular} & \begin{tabular}[c]{@{}c@{}}-97 dBm/\\ -76 dBm\end{tabular} \\ \hline
TV & Lenovo & Ideatv 55E31Y & \begin{tabular}[c]{@{}c@{}}-77 dBm/\\ -78 dBm\end{tabular} & \begin{tabular}[c]{@{}c@{}}-93 dBm/\\ -74 dBm\end{tabular} \\ \hline
\multicolumn{5}{c}{This test used VGA (1.5m) and HDMI1.4 (1.5m) made by UGREEN.}\\ \hline
\end{tabular}
}
\label{Tab:Cross-Device}
\end{table}

Next, we test \ourprotocol on different display devices. Detailed manufacturers and models are shown in Table~\ref{Tab:Cross-Device}. The results show that \ourprotocol shows a wide range of feasibility on display devices from different manufacturers/models, and there is no significant difference in the RSSI values returned by COTS LoRa receivers. Additionally, although the main attack scenario of \ourprotocol is a monitor connected to a video cable, it can also work on projectors and TVs.

\begin{figure}[t]
\centering
\includegraphics[width=0.98\linewidth]{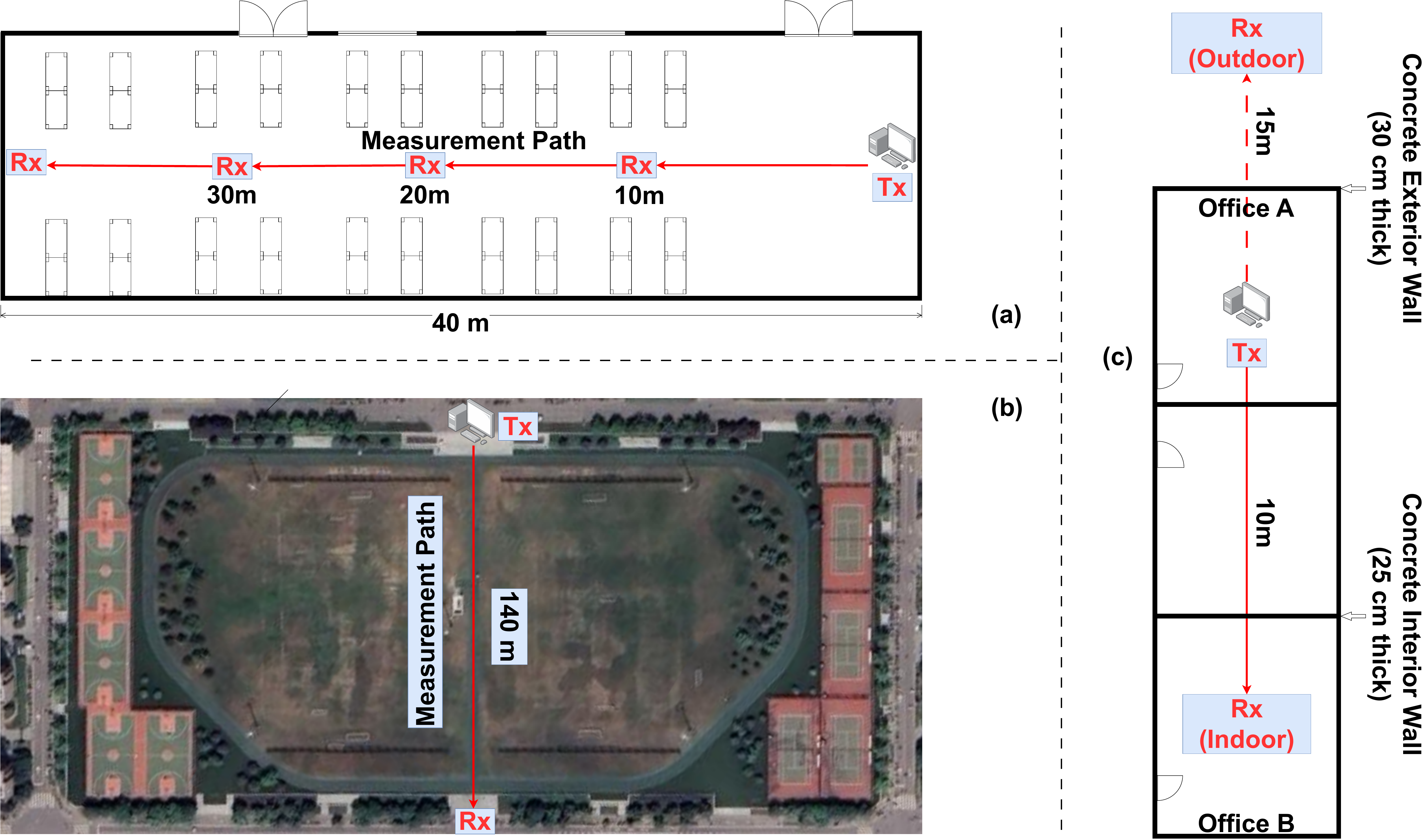}
\caption{Indoor, outdoor, and through-wall scenes.}
\label{fig:Map}
\end{figure}

\subsection{Attack Distance}
We measure the maximum attack distance both indoors and outdoors. The attack video loops on the screen (\ie, the video cable repeatedly emits EM packets), and the EM packet's payload length is fixed to 24 raw bits. In this section, the 'maximum attack distance' is defined as the farthest distance at which the receiver can reliably decode the EM packets. This distance is determined by moving Rx along the Measurement Path in Figure~\ref{fig:Map} (a) and (b) until the receiver is no longer able to receive and decode EM packets completely and correctly. Around the extreme attack distance, the receiver may detect LoRa packets, but the decoded Payload contains errors Due to signal attenuation. Therefore, the maximum attack distance we record is the farthest location where correctly decoded EM packets exist. For the evaluation of COTS LoRa, we ensure that the SF and BW of the EM packets match the settings of the LoRa receivers with SF ranging from 6 to 12 and BW options at 125 kHz, 250 kHz, and 500 kHz. 

\begin{table}[b]
\caption{Indoor attack distances on COTS LoRa.}
\resizebox{1\columnwidth}{!}{
\begin{tabular}{c|ccccccc}
\hline                               
\textbf{433MHz\&VGA} & SF6   & SF7   & SF8   & SF9   & SF10  & SF11  & SF12 \\
\hline
125kHz & 36.7m & 40m   & 40m   & 40m   & 40m   & 8.8m  & 1.0m \\
250kHz & 28.5m & 32m   & 40m   & 40m   & 40m   & 35.2m & 3.2m \\
500kHz & 23.0m & 17.9m & 30.4m & 40m   & 40m   &  40m   & 6.5m \\
\hline 
\textbf{433MHz\&HDMI}    & SF6   & SF7   & SF8   & SF9   & SF10  & SF11  & SF12 \\
\hline 
125kHz & 30.4m & 37.2m &  40m   & 40m   & 40m   & 6.0m    & 1.0m \\
250kHz & 28.7m & 28.7m & 37.5m & 40m   & 40m   & 32m   & 3.5m \\
500kHz & 17.2m & 16.6m & 30.9m & 40m   & 40m   & 40m   & 6.0m \\
\hline 
\textbf{915MHz\&VGA}     & SF6   & SF7   & SF8   & SF9   & SF10  & SF11  & SF12 \\
\hline 
125kHz & 25.0m & 25.2m & 26.5m & 33.5m & 30.5m & 8.5m  & 1.0m \\
250kHz & 22.0m & 22.5m & 26.2m & 30.0m & 27.4m & 19.1m & 3.4m \\
500kHz & 17.2m & 17.9m & 22.5m & 27.7m & 26.5m & 23.5m & 5.3m \\
\hline
\textbf{915MHz\&HDMI} & SF6   & SF7   & SF8   & SF9   & SF10  & SF11  & SF12 \\
\hline 
125kHz & 40m   & 40m   & 40m   & 40m   & 40m   & 6.0m  & 1.0m \\
250kHz & 40m   & 40m   & 40m   & 40m   & 40m   & 19.5m & 3.5m \\
500kHz & 40m   & 40m   & 40m   & 40m   & 40m   & 40m   & 6.5m \\
\hline 
\end{tabular}
}
\label{TAB:IndoorDistance}
\end{table}

\begin{figure}[t]
\centering
\includegraphics[width=0.7\linewidth]{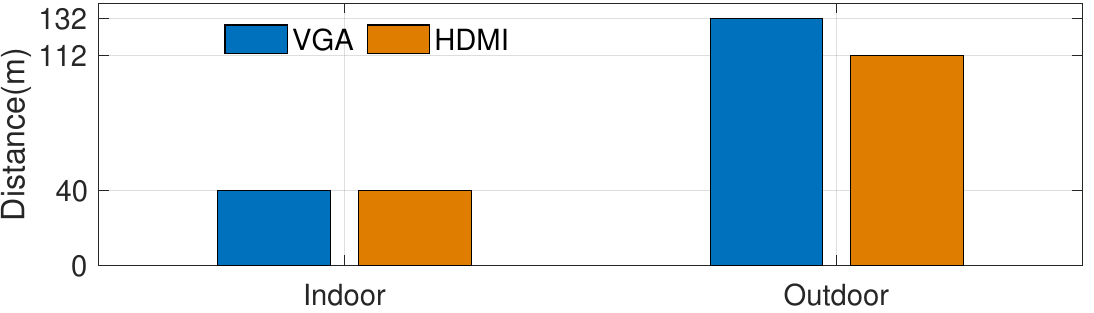}
\caption{Maximum attack distances on SDR.}
\label{fig:SDRDistance}
\end{figure}

\textbf{Indoor}: The indoor testing is conducted in an office as shown in Figure~\ref{fig:Map} (a), where the distance between the two endpoints is 40m, a distance we deem sufficient to exceed the lengths of most real-world offices. The victim's computer (Tx) is placed on the far right side of the room. The receiver moves along the measurement path until the EM packet can no longer be decoded correctly. Table~\ref{TAB:IndoorDistance} shows the maximum attack distances in the office scene. 
As the chirp duration (starting from SF6\&500 kHz) increases, the attack distance gradually increases until it reaches 40m at the end of the room. This combination of SF\&BW and the growth relationship of transmission distance are consistent with LoRa technology. At 433 MHz, between SF7 and SF11, both VGA and HDMI can support a transmission distance of up to 40m. At 915 MHz, VGA's maximum attack distance is 33.5m at SF8\&125 kHz, while HDMI's maximum attack distance can be 40m. However, when the chirp duration surpasses the threshold of SF11\&250 kHz, the attack distance drops sharply. This is because as the chirp duration increases, the number of frames required for the attack videos increases, resulting in increased errors due to the frame intervals.
Figure~\ref{fig:SDRDistance} shows the maximum attack distance of SDR-based version.
With the log-periodic antenna (higher receiving gain) and higher leakage intensity at $f_{sdr}$ frequency, \ourprotocol can achieve the maximum attack distance of 40m. 

\begin{table}[t]
\caption{Outdoor attack distances on COTS LoRa.}
\resizebox{1\columnwidth}{!}{
\begin{tabular}{c|ccccccc} 
\hline
\textbf{433MHz\&VGA}       & SF6   & SF7   & SF8                          & SF9                          & SF10  & SF11                        & SF12  \\
\hline
125kHz & 40.1m & 46.0m & 47.4m & \textbf{52.2m} & 50.5m & 8.5m & 1.5m  \\
250kHz & 27.6m & 38.5m & 42.0m                        & 47.5m                        & 47.1m & 26.0m                      & 3.5m  \\
500kHz & 23.4m & 30.6m & 37.6m                        & 43.5m                        & 41.9m & 43.0m                       & 6.5m  \\
\hline
\textbf{433MHz\&HDMI}       & SF6   & SF7   & SF8                          & SF9                          & SF10  & SF11                        & SF12  \\
\hline
125kHz & 36.6m & 45.0m & 46.2m                        & 50.0m                        & \textbf{51.0m} & 12.1m                       & 1.0m  \\
250kHz & 31.9m & 37.1m & 42.0m                        & 44.4m                        & 47.5m & 19.6m                       & 3.0m  \\
500kHz & 23.4m & 29.4m & 36.5m                        & 41.4m                        & 42.1m & 39.5m                       & 4.0m  \\
\hline                                                            
\textbf{915MHz\&VGA}       & SF6   & SF7   & SF8                          & SF9                          & SF10  & SF11                        & SF12  \\
\hline
125kHz & 23.1m & 27.8m & 32.0m                        & \textbf{39.5m}                        & 36.5m & 10.7m                       & 1.0m  \\
250kHz & 20.1m & 22.3m & 29.5m                        & 36.7m                        & 34.2m & 17.5m                       & 3.3m  \\
500kHz & 15.4m & 19.7m & 24.0m                        & 28.5m                        & 30.5m & 29.4m                       & 6.0m  \\
\hline
\textbf{915MHz\&HDMI} & SF6   & SF7   & SF8                          & SF9                          & SF10  & SF11                        & SF12  \\
\hline
125kHz & 64.2m & 68.4m & 73.5m                        & \textbf{87.5m}                       & 87.2m & 17.5m                       & 3.5m  \\
250kHz & 61.9m & 67.1m & 70.5m                        & 77.4m                        & 83.5m & 24.3m                       & 7.5m  \\
500kHz & 53.1m & 61.0m & 66.1m                        & 72.8m                        & 69.2m & 50.0m                       & 10.0m \\
\hline
\end{tabular}
}
\label{TAB:OutdoorDistance}
\end{table}

\textbf{Outdoor}: For the outdoor scenario, we evaluate the maximum attack distance on a playground shown in Figure~\ref{fig:Map} (b). The receiver is moved along the measurement path from the Tx to determine the maximum attack distance. Table~\ref{TAB:OutdoorDistance} shows the maximum attack distances supported by COTS LoRa in outdoor environments. The outdoor evaluation mirrors the indoor evaluation, with the attack distance incrementally extending as the chirp duration until SF11\&250 kHz. Beyond this threshold, the attack distance decreases due to the frame interval.
At 433 MHz, the maximum distance is 52.2m at SF9\&125 kHz with the VGA cable, and 51.0m at SF10\&125 kHz for HDMI. In the 915 MHz band, the HDMI cable performs better, and the attack distance that can be achieved at the SF9\&125 kHz settings is 87.5m. While using VGA cable at SF9\&125 kHz, the distance reaches 39.5m. When using an SDR-based receiver, the maximum distance is extended to 112m (HDMI) - 132m (VGA).

Comparing the results indoors and outdoors, the absence of obstacles outdoors typically results in longer attack distances compared to indoor environments. 
On the non-LoRa frequency bands, the higher emission intensity and higher antenna gain enable attackers to capture secret data from over hundred-meter away using the SDR. 
In addition, we notice that when the victim uses an HDMI cable, the attack distance at 915 MHz (87.5m) is longer than that at 433 MHz (51m), although the leakage intensity of 433 MHz (47.8 dB) is slightly higher than that of 915 MHz (43.1 dB) as shown in Figure~\ref{fig:LeakSweep}. 
We speculate that the main reason for this difference is that the sensitivity of LoRa gateways is higher than that of individual LoRa nodes.
From the attacker's perspective, the malware can initially identify the victim's cable type. If it is VGA, it could emit EM packets at 433 MHz; if it is HDMI, 915 MHz is more effective. 
Subsequently, the malware can select the appropriate settings (such as SF9\&125 kHz, or use the SDR-based version of \ourprotocol) to decode secret data at a longer distance.

\subsection{Through-wall Transmission}
Next, we evaluate the through-wall transmission in the scenarios shown in Figure~\ref{fig:Map} (c). 
The Tx is located in Office A. The distance between Tx and the location of the outdoor receiver (Rx) is 15m, obstructed by a 30cm-thick concrete exterior wall; the distance between Tx and indoor Rx is 10m, separated by two concrete interior walls (25cm thick). 
To evaluate the receiving capability of COTS LoRa, Tx emits EM packets at 433 MHz via VGA cable and at 915 MHz via the HDMI cable with SF9\&125 kHz settings (the combination for the optimal attack distance); we record the RSSI value returned by the LoRa node/gateway. For the SDR-based version, we compute the DNR of captured signals from VGA and HDMI.

\begin{table}[h]
\caption{The EM packets' RSSI (for COTS LoRa) and DNR (for SDR) after through-wall transmission.}
\resizebox{0.85\columnwidth}{!}{
\begin{tabular}{c|c|c|c}
\hline
\begin{tabular}[c]{@{}c@{}}Through wall\end{tabular}       & 433MHz                                                     & 915MHz                                                  & \begin{tabular}[c]{@{}c@{}} SDR-based \end{tabular} \\ \hline
\begin{tabular}[c]{@{}c@{}}One Exterior\\  Wall\end{tabular}  & \begin{tabular}[c]{@{}c@{}}-108 dBm -\\ -112 dBm\end{tabular} & \begin{tabular}[c]{@{}c@{}}-108dBm -\\ -111 dBm\end{tabular}  & \begin{tabular}[c]{@{}c@{}}18.7 (VGA)\\14.7 (HDMI)\end{tabular} \\ \hline
\begin{tabular}[c]{@{}c@{}}Two Interior \\ Walls\end{tabular} & \begin{tabular}[c]{@{}c@{}}-116 dBm -\\ -120 dBm\end{tabular}  & \begin{tabular}[c]{@{}c@{}}-115 dBm -\\ -118 dBm\end{tabular} & \begin{tabular}[c]{@{}c@{}}11.6 (VGA)\\7.4 (HDMI)\end{tabular}   \\ \hline
\end{tabular}
}
\label{Tab:ThroughWall}
\end{table}

Table~\ref{Tab:ThroughWall} shows the RSSI and DNR values for through-wall transmission using COTS LoRa devices and SDR. 
In the 'indoor to outdoor' scenario, the RSSI values reported range from -108 dBm to -112 dBm. While in 'indoor to indoor', transmissions experience more significant attenuation after passing through two interior walls, with RSSI decreasing from -115 dBm to -120 dBm. As a comparison, the minimum RSSI we observed in attack distance evaluation is between -120 dBm and -124 dBm. 
For the SDR-based version, we first measure the initial DNR of chirps at 1m away from Tx, which is 31.9 for VGA and 24.3 for HDMI. After penetrating one exterior wall, the DNR of VGA and HDMI drops to 18.7 and to 14.7, respectively; while after penetrating two interior walls, EM emission is further attenuated, with the DNR of VGA dropping to 11.6 and that of HDMI dropping to 7.4.

\begin{figure}[t]
    \begin{minipage}{0.99\linewidth}
        \subfigure[]{
                \centering
        \includegraphics[width=0.47\linewidth]{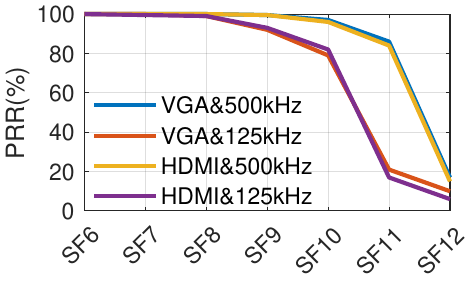}
        \label{fig:PRRCombination}
        }
        \subfigure[]{
        \centering
	\includegraphics[width=0.47\linewidth]{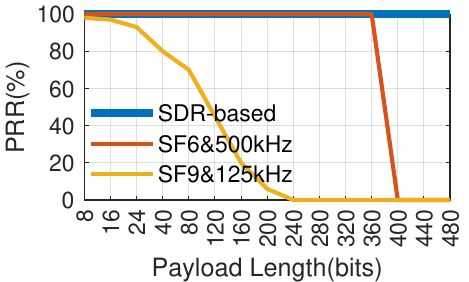}
	\label{fig:PRRPayloadLength}
        }
        \caption{(a) PRR under various SF\&BW. (b) PRR under various payload lengths.}
    \end{minipage}
\end{figure}

\subsection{Packet Reception Rate}
\label{Sec:PRR}
To quantify the impact of signal intervals on EM packets, we measure the packet reception rate (PRR) in the office scenarios shown in Figure~\ref{fig:Map} (a). The EM packets were emitted 1000 times; successful packet reception is defined as all bits in the EM packet being decoded correctly. We evaluate 433 MHz using VGA and 915 MHz using HDMI.

\textbf{Combination of SF and BW}: The settings include SF6 to SF12, paired with BWs of 125 kHz and 500 kHz, and the payload length is 24 raw bits. The COTS LoRa receivers are placed at half the maximum attack distance indoors as shown in Table~\ref{TAB:IndoorDistance} (\eg, 16.75m from the monitor when at the HDMI\&SF9\&125 kHz setting).

As shown in Figure~\ref{fig:PRRCombination}, the PRR exhibits a gradual decline from 100\% to around 80\% as the SF and BW increase and drop sharply when the SF is higher than 11 at BW of 500 kHz. In contrast, this turning point is at SF10 if the BW is 125 kHz.
The reason is that the longer chirp duration requires more frames of attack video, which increases the possibility of EM packets being decoded incorrectly. Notably, under the same SF\&BW settings, the cable type has little impact on the PRR.

\textbf{Payload Length}: Given the minimal impact of cable type on the PRR, in this evaluation, we use HDMI at 915 MHz to measure the PRR across payload lengths ranging from 8 raw bits to 480 raw bits. We select two representative combinations: SF6\&500 kHz (with the minimum chirp duration) and SF9\&125 kHz (with the longest attack distance). The COTS LoRa receiver/SDR is positioned 20m/40m from Tx.

As shown in Figure~\ref{fig:PRRPayloadLength}, under SF6\&500 kHz, the PRR sustains at 100\% when the payload length is less than or equal to 360 bits. However, it plummets to 0\% when the payload length reaches 400 bits. This decline occurs because the 400-bit payload needs about 1.02 frames (that is, the attack video containing 2 frames), and one frame interval could destroy 5.2 chirps at SF6\&500 kHz, such broken packets exceed LoRa's tolerance. Conversely, under the SF9\&125 kHz setting, the PRR is 93\% for 24-bit payload and begins to decline sharply beyond 40 bits. The duration of the frame interval is roughly 2\% of a single chirp, which suggests a lower likelihood of EM packet corruption. Nonetheless, SF9\&125 kHz requires more frames under the same payload length, and the PRR starts to drop at a shorter payload length. On the SDR, benefiting from the design of aligning signal interval, the EM packets with payload lengths ranging from 8 bits to 480 bits can be successfully decoded at a distance of 40m.

\begin{figure}
    \centering
    \includegraphics[width=0.95\linewidth]{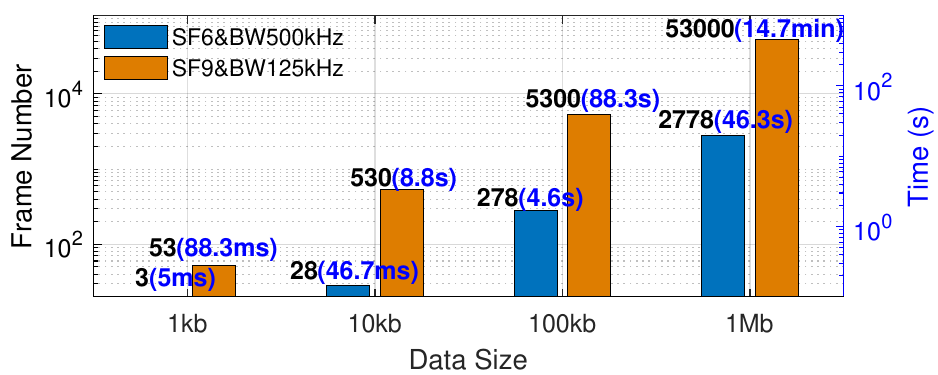}
    \caption{The frame number of attack video and the playback time for transmitting different amounts of secret data.} 
    \label{fig:SizeFrameNum}
\end{figure}

\subsection{Data Rate}
\textbf{Data size}: As shown in Figure~\ref{fig:SizeFrameNum}, we selected the two most representative settings of SF6\&BW500 kHz and SF9\&BW125kHz to evaluate the number of attack video frames and the corresponding video time. For a short message of 1 kb (\eg, access keys), \ourprotocol can complete the transmission in less than 1 second. For confidential information of 10 kb to 100 kb (e.g., log files), it only takes 46.7ms to 4.6s under the fastest setting SF6\&BW500 kHz, and SF9\&BW125 kHz takes between 8.8s and 88.3s. 

\textbf{Goodput}: Next, we evaluate the actual throughput (goodput, the actual data rate after the error packets are removed) in indoors, including SF6\&500 kHz, SF9\&125 kHz, and SDR-based version, at frequencies 433 MHz (VGA), 915 MHz (HDMI), and $f_{sdr}$. The theoretical throughputs corresponding to these settings are 21.6 kbps, 1160 bps (which are equal to the standard LoRa's throughput), and 180 bps, respectively. The receiver is positioned at distances of 10m, 20m, and 40m from Tx. During each test, Tx emits 1000 EM packets, and we calculate the goodput.

\begin{table}[h]
\caption{Goodput at 10m, 20, 30m, and 40m.}
\resizebox{1\columnwidth}{!}{
\begin{tabular}{c|ccc}
\hline
\textbf{433MHz\&VGA}  & SF6\&500kHz & SF9\&125kHz & SDR-based \\
\hline
10m         & 21.6 kbps   & 1160 bps     & 180 bps   \\
20m         & 20.5 kbps   & 1156.5 bps     & 180 bps   \\
30m         & /           & 1153 bps     & 179.6 bps \\
40m         & /           & 1148.4 bps   & 179.2 bps   \\
\hline
\textbf{915MHz\&HDMI} & SF6\&500kHz & SF9\&125kHz & SDR-based \\
\hline
10m         & 21.6 kbps   & 1160 bps     & 180 bps   \\
20m         & 21.53 kbps   & 1160 bps     & 180 bps   \\
30m         & 21.38 kbps  & 1154.2 bps     & 178.9 bps  \\
40m         & 21.04 kbps  & 1151.9 bps   & 178.4 bps  \\
\hline
\end{tabular}
}
\label{Tab:Goodput}
\end{table}

Table~\ref{Tab:Goodput} shows the goodput in the three settings at 10m to 40m. At 10m and 20m, \ourprotocol only has slight packet loss, and the goodput is equal to or close to the theoretical maximum throughput. As the distance increases to 30m to 40m, except for SF6\&500kHz\&VGA setting, which is not measured due to signal attenuation, the goodput of other settings decreases slightly. In general, in an office scenario, attackers can choose the SF6\&BW500 kHz to transmit sensitive data at the fastest speed; if the receiver is far away from the victim, using SF9\&BW125 kHz setting or the SDR-based version can capture secret data at a slightly lower rate but a longer distance.

\subsection{Impact of Receiving Direction}
We measure the emission intensity in various receiving directions for four potential cable placements: vertical, horizontal, circular, and curved. The Rx is placed at a distance of 10 meters from the monitor at different directions to record the RSSI values.

\begin{figure}[h]
    \begin{minipage}{0.93\linewidth}
        \subfigure{
                \centering
        \includegraphics[width=0.95\linewidth]{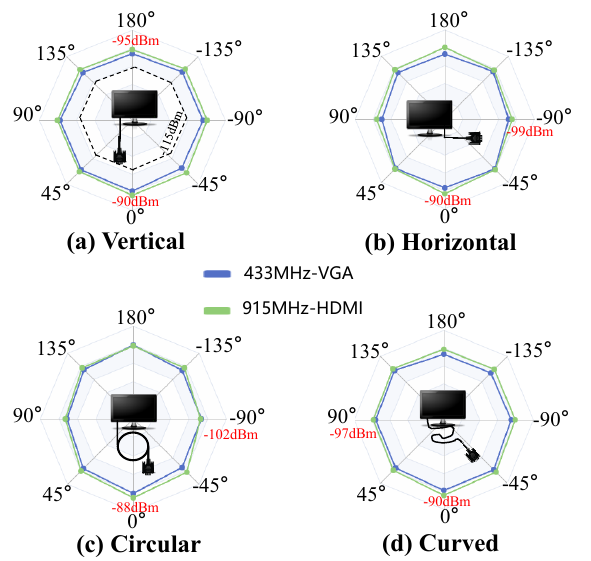}
        \label{fig:EmissionDirection}
        }
        \caption{RSSI values in various directions.} 
    \end{minipage} 
\end{figure}

As shown in Figure~\ref{fig:EmissionDirection}, overall, the RSSI of HDMI at the 915 MHz attack frequency is slightly higher than that of VGA at 433 MHz, and we marked the maximum and minimum RSSI of 915 MHz\&HDMI in the corresponding direction. In all placement methods, the RSSI in front of the monitor (0°) is the highest (-88 dBm to -90 dBm). In the vertical placement, the position of minimum RSSI (-95 dBm) is behind the monitor (180°) because the monitor itself blocks part of the leakage. When placed horizontally, the RSSI on the sides of the monitor (90° and -90°) is the smallest (about -99 dBm). If the cable is placed casually, RSSIs range from -90 dBm to -97 dBm. Overall, the EMR intensity in different directions is relatively uniform, implying that attackers are not restricted to specific orientations when receiving the secret packets.

\section{Discussion}
In this section, we discuss considerations for practical deployment, \ourprotocol future work and countermeasures.

\textbf{Multiple active monitors}: The malware carrying \ourprotocol may infect multiple computers in proximity to the victim. If these active monitors simultaneously emit EM packets using the same LoRa settings, the receiving end may have difficulty distinguishing them. A straightforward solution is to embed a unique hardware identifier (\eg, CPU ID or motherboard UUID, which is accessible at the software level) in each EM packet to enable source differentiation at the receiver. Furthermore, malware can utilize these hardware IDs to schedule transmission at randomized times, reducing the likelihood of packet collisions.

\textbf{\ourprotocol and CTCC's future extension}: Although this paper uses a common display setting of 1080x1920@60Hz for analysis and evaluation, \ourprotocol can be adapted to other resolutions and refresh rates by adjusting the input parameters ($ScreenH$, $ScreenW$, and $PC$ under the corresponding display setting) in Algorithm~\ref{Alg:EMChirp}, enabling COTS LoRa devices to decode the EM packets.

Additionally, on the basis of our experimental observations, we believe that the concept of CTCC can be extended to other commercial wireless technologies. First, we observe that the video cable can emit EMR beyond 1000 MHz as discussed in Figure~\ref{fig:LeakSweep}. For example, HDMI cables still exhibit measurable EMR intensity at 2.4 GHz (WiFi and Bluetooth's frequency bands). Second, while generating waveforms compatible with other wireless protocols requires more sophisticated modulation techniques, it is feasible. For instance, EMR's amplitude can be more finely controlled by changing the grayscale value of the pixel (under the commonly used 8-bit RGB setting, this provides up to 256 levels of intensity control, allowing the video cable EMR transmitter to approximate 8-bit amplitude resolution). Furthermore, the EMR phase can also be manipulated by controlling the spatial arrangement of black-white pixels.

\textbf{Countermeasures Analysis}: Given the non-privilege, high flexibility of attack frequencies, and strong attenuation resistance of \ourprotocol, traditional countermeasures against physical covert channels need to be re-examined and improved:

(1) EM shielding. The video cables we experimented with in the evaluation are shielded with aluminum foil, copper braid, and twisted pair designs to reduce EMR. These shielding methods could protect against conventional EM covert channels, but fail to defend against attackers armed with \ourprotocol. The main reason is that the LoRa-compatible EM packets could benefit from strong noise resilience of LoRa wireless technology and high sensitivity of LoRa radios and be received at much greater distances even when the video cables are shielded and EM signals are minimized. 
We call for innovative shielding methods to mitigate the risks posed by CTCC techniques.

(2) RF jammers. RF jammers could overpower the leaked signal with high-power noise and obstruct CTCC's reception. This potential protection method would inevitably disrupt normal LoRa communications widely deployed worldwide as well as other coexisting wireless communications in the ISM bands such as WiFI and wireless radios for medical devices, which makes these countermeasures infeasible in practice.  
Moreover, attackers could adaptively select different parameter configurations such as SF, BW and even central frequency to evade from RF jammers and covert channel detection. 

(3) Covert packet detection. Unlike prior EM covert channels that rely on custom modulation schemes with distinctive spectral patterns (such as B-FSK), the LoRa-compatible waveforms making covert EM packets resemble legitimate LoRa transmissions on the spectrum. This similarity allows them to bypass conventional RF anomaly detection based on spectral signatures. However, subtle waveform differences still exist between LoRa-like EMR signals and genuine LoRa transmission. These differences do not affect packet decoding, but may serve as a basis for countermeasures. Future work could explore detection algorithms based on physical-layer fingerprints to differentiate covert EM packets from legitimate ones.

(4) Larger isolation area. Given \ourprotocol's over-hundred-meter attack range, the isolation areas for conventional air-gapped networks must be further expanded. Eliminating all COTS wireless devices from an entire building or creating a Faraday zoom is viable but incurs prohibitive implementation cost. Considering the high penetration of LoRa technologies supporting a variety of IoT applications worldwide, it is increasingly likely that attackers could self-deploy or leverage existing third-party LoRa infrastructure near a target to receive LoRa-compatible EM packets.  
Currently, only a few highly confidential scenarios (\eg, military applications~\cite{MilitaryShielding}) can afford and build large shielding areas while other sectors (\eg, finance, business, medical, \etc) remain vulnerable to the risk of this new type of covert channel attacks reported in this paper. 

\section{Conclusion}
This paper reveals the risk of covertly leaking sensitive information from air-gapped computers by generating LoRa-compatible EM packets with video cables. The EM packets can be received from widely deployed COTS LoRa devices from afar, even when the video cables are protected behind concrete walls. Attackers also could go beyond and break the limit of the communication range of  the LoRa protocol by crafting customized EM packets and receive with low-cost SDRs. The presented CTCC technologies with fine-grained control of EM leakage over a large wireless spectrum can potentially be applied to generate EM packets compatible with other wireless technologies such as WiFi and ZigBee \footnote{Preliminary results obtained but not reported in this paper due to page limit}. We plan to comprehensively investigate such potential risks and possible countermeasures in the future. 

\section{Open Source}
To promote reproducibility and further research, we have released the full implementation of \ourprotocol as open-source and permanently archived it on Zenodo: \url{https://zenodo.org/records/15532223}. We provide comprehensive reproduction materials, including the source code, detailed configuration settings for the monitor, the LoRa SX1262 node, and the SX1302 gateway, as well as some artifacts such as attack images generated under varying attack frequencies, spreading factors (SF), and bandwidths (BW).

\begin{acks}
We thank the anonymous reviewers for their insightful comments and constructive suggestions, which helped improve the quality of this paper. 
This work was supported by the National Key R\&D Program of China 2023YFB2904000, the NSFC Grant No. 62302383, as well as the Hong Kong GRF Grant No. 15211924 and 15206123.
\end{acks}

\bibliographystyle{ACM-Reference-Format}
\balance
\bibliography{sample-base}


\begin{thebibliography}{76}


\ifx \showCODEN    \undefined \def \showCODEN     #1{\unskip}     \fi
\ifx \showISBNx    \undefined \def \showISBNx     #1{\unskip}     \fi
\ifx \showISBNxiii \undefined \def \showISBNxiii  #1{\unskip}     \fi
\ifx \showISSN     \undefined \def \showISSN      #1{\unskip}     \fi
\ifx \showLCCN     \undefined \def \showLCCN      #1{\unskip}     \fi
\ifx \shownote     \undefined \def \shownote      #1{#1}          \fi
\ifx \showarticletitle \undefined \def \showarticletitle #1{#1}   \fi
\ifx \showURL      \undefined \def \showURL       {\relax}        \fi
\providecommand\bibfield[2]{#2}
\providecommand\bibinfo[2]{#2}
\providecommand\natexlab[1]{#1}
\providecommand\showeprint[2][]{arXiv:#2}

\bibitem[Arendt(2016)]%
        {arendt2016medical}
\bibfield{author}{\bibinfo{person}{Daniel Arendt}.} \bibinfo{year}{2016}\natexlab{}.
\newblock \showarticletitle{Medical-Grade Network Security-Air-Gap Isolation and PossibleWeak Points}.
\newblock \bibinfo{journal}{\emph{Journal of Applied Computer Science}} \bibinfo{volume}{24}, \bibinfo{number}{3} (\bibinfo{year}{2016}), \bibinfo{pages}{7--19}.
\newblock


\bibitem[Association(2013)]%
        {VESA}
\bibfield{author}{\bibinfo{person}{Video Electronics~Standards Association}.} \bibinfo{year}{2013}\natexlab{}.
\newblock \bibinfo{title}{Video Electronics Standards Association, Display Monitor Timing 1.3}.
\newblock \bibinfo{howpublished}{\url{https://glenwing.github.io/docs/VESA-DMT-1.13.pdf}}.
\newblock


\bibitem[Barker and Barker(2018)]%
        {barker2018recommendation}
\bibfield{author}{\bibinfo{person}{Elaine Barker} {and} \bibinfo{person}{William Barker}.} \bibinfo{year}{2018}\natexlab{}.
\newblock \bibinfo{booktitle}{\emph{Recommendation for key management, part 2: best practices for key management organization}}.
\newblock \bibinfo{type}{{T}echnical {R}eport}. \bibinfo{institution}{National Institute of Standards and Technology}.
\newblock


\bibitem[Boichat(2006)]%
        {DDCcontrol}
\bibfield{author}{\bibinfo{person}{Nicolas Boichat}.} \bibinfo{year}{2006}\natexlab{}.
\newblock \showarticletitle{DDCcontrol documentation}.
\newblock \bibinfo{journal}{\emph{[Online] https://ddccontrol.sourceforge.net/doc/ddccontrol-0.4.pdf}} (\bibinfo{year}{2006}).
\newblock


\bibitem[Camurati and Francillon(2022)]%
        {Noise-SDR}
\bibfield{author}{\bibinfo{person}{Giovanni Camurati} {and} \bibinfo{person}{Aur{\'e}lien Francillon}.} \bibinfo{year}{2022}\natexlab{}.
\newblock \showarticletitle{Noise-SDR: Arbitrary Modulation of Electromagnetic Noise from Unprivileged Software and Its Impact on Emission Security}. In \bibinfo{booktitle}{\emph{2022 IEEE Symposium on Security and Privacy (SP)}}. IEEE, \bibinfo{pages}{1193--1210}.
\newblock


\bibitem[Cho and Shin(2021)]%
        {BlueFi}
\bibfield{author}{\bibinfo{person}{Hsun-Wei Cho} {and} \bibinfo{person}{Kang~G Shin}.} \bibinfo{year}{2021}\natexlab{}.
\newblock \showarticletitle{BlueFi: bluetooth over WiFi}. In \bibinfo{booktitle}{\emph{Proceedings of the 2021 ACM SIGCOMM 2021 Conference}}. \bibinfo{pages}{475--487}.
\newblock


\bibitem[Chua(2005)]%
        {chua2005cyberciege}
\bibfield{author}{\bibinfo{person}{Chay Chua}.} \bibinfo{year}{2005}\natexlab{}.
\newblock \emph{\bibinfo{title}{CyberCIEGE scenario illustrating software integrity issues and management of air-gapped networks in a military environment}}.
\newblock \bibinfo{thesistype}{Ph.\,D. Dissertation}. \bibinfo{school}{Monterey, California. Naval Postgraduate School}.
\newblock


\bibitem[Cloud(2024)]%
        {GoogleStd}
\bibfield{author}{\bibinfo{person}{Google Cloud}.} \bibinfo{year}{2024}\natexlab{}.
\newblock \bibinfo{booktitle}{\emph{How Google protects the physical-to-logical space in a data center [online]}}.
\newblock
\urldef\tempurl%
\url{https://cloud.google.com/docs/security/physical-to-logical-space}
\showURL{%
\tempurl}


\bibitem[Deng et~al\mbox{.}(2020)]%
        {deng2020novel}
\bibfield{author}{\bibinfo{person}{Fangming Deng}, \bibinfo{person}{Pengqi Zuo}, \bibinfo{person}{Kaiyun Wen}, {and} \bibinfo{person}{Xiang Wu}.} \bibinfo{year}{2020}\natexlab{}.
\newblock \showarticletitle{Novel soil environment monitoring system based on RFID sensor and LoRa}.
\newblock \bibinfo{journal}{\emph{Computers and Electronics in Agriculture}}  \bibinfo{volume}{169} (\bibinfo{year}{2020}), \bibinfo{pages}{105169}.
\newblock


\bibitem[et~al.(2006)]%
        {HDMIStd}
\bibfield{author}{\bibinfo{person}{LLC et al.}} \bibinfo{year}{2006}\natexlab{}.
\newblock \bibinfo{title}{,“High-Definition Multimedia Interface Specification Version 1.3 a: Supplement 1 Consumer Electronics Control (CEC)”}.
\newblock


\bibitem[Feng et~al\mbox{.}(2023)]%
        {SideComm}
\bibfield{author}{\bibinfo{person}{Justin Feng}, \bibinfo{person}{Timothy Jacques}, \bibinfo{person}{Omid Abari}, {and} \bibinfo{person}{Nader Sehatbakhsh}.} \bibinfo{year}{2023}\natexlab{}.
\newblock \showarticletitle{Everything has its Bad Side and Good Side: Turning Processors to Low Overhead Radios Using Side-Channels}. In \bibinfo{booktitle}{\emph{Proceedings of the 22nd International Conference on Information Processing in Sensor Networks}}. \bibinfo{pages}{288--301}.
\newblock


\bibitem[FireEye(2020)]%
        {fireeye2020highly}
\bibfield{author}{\bibinfo{person}{FireEye}.} \bibinfo{year}{2020}\natexlab{}.
\newblock \showarticletitle{Highly evasive attacker leverages SolarWinds supply chain to compromise multiple global victims with SUNBURST backdoor}.
\newblock \bibinfo{journal}{\emph{FireEye Threat Research}} (\bibinfo{year}{2020}).
\newblock


\bibitem[Force(2017)]%
        {force2017security}
\bibfield{author}{\bibinfo{person}{Joint~Task Force}.} \bibinfo{year}{2017}\natexlab{}.
\newblock \bibinfo{booktitle}{\emph{Security and privacy controls for information systems and organizations}}.
\newblock \bibinfo{type}{{T}echnical {R}eport}. \bibinfo{institution}{National Institute of Standards and Technology}.
\newblock


\bibitem[Gadgets(2018)]%
        {HackRF}
\bibfield{author}{\bibinfo{person}{Great~Scott Gadgets}.} \bibinfo{year}{2018}\natexlab{}.
\newblock \bibinfo{booktitle}{\emph{Hackrf one Official Website [online]}}.
\newblock
\urldef\tempurl%
\url{https://greatscottgadgets.com/hackrf}
\showURL{%
\tempurl}


\bibitem[Gaw{\l}owicz et~al\mbox{.}(2022)]%
        {Wi-Lo}
\bibfield{author}{\bibinfo{person}{Piotr Gaw{\l}owicz}, \bibinfo{person}{Anatolij Zubow}, {and} \bibinfo{person}{Falko Dressler}.} \bibinfo{year}{2022}\natexlab{}.
\newblock \showarticletitle{Wi-Lo: Emulation of LoRa using Commodity 802.11 b WiFi Devices}. In \bibinfo{booktitle}{\emph{ICC 2022-IEEE International Conference on Communications}}. IEEE, \bibinfo{pages}{4414--4419}.
\newblock


\bibitem[Gu and Peng(2010)]%
        {GSMProtocol}
\bibfield{author}{\bibinfo{person}{Guifen Gu} {and} \bibinfo{person}{Guili Peng}.} \bibinfo{year}{2010}\natexlab{}.
\newblock \showarticletitle{The survey of GSM wireless communication system}. In \bibinfo{booktitle}{\emph{2010 international conference on computer and information application}}. IEEE, \bibinfo{pages}{121--124}.
\newblock


\bibitem[Guri(2022)]%
        {Air-Fi}
\bibfield{author}{\bibinfo{person}{Mordechai Guri}.} \bibinfo{year}{2022}\natexlab{}.
\newblock \showarticletitle{Air-fi: Leaking data from air-gapped computers using wi-fi frequencies}.
\newblock \bibinfo{journal}{\emph{IEEE Transactions on Dependable and Secure Computing}} (\bibinfo{year}{2022}).
\newblock


\bibitem[Guri and Elovici(2018)]%
        {Bridgeware}
\bibfield{author}{\bibinfo{person}{Mordechai Guri} {and} \bibinfo{person}{Yuval Elovici}.} \bibinfo{year}{2018}\natexlab{}.
\newblock \showarticletitle{Bridgeware: The air-gap malware}.
\newblock \bibinfo{journal}{\emph{Commun. ACM}} \bibinfo{volume}{61}, \bibinfo{number}{4} (\bibinfo{year}{2018}), \bibinfo{pages}{74--82}.
\newblock


\bibitem[Guri et~al\mbox{.}(2015)]%
        {GSMem}
\bibfield{author}{\bibinfo{person}{Mordechai Guri}, \bibinfo{person}{Assaf Kachlon}, \bibinfo{person}{Ofer Hasson}, \bibinfo{person}{Gabi Kedma}, \bibinfo{person}{Yisroel Mirsky}, {and} \bibinfo{person}{Yuval Elovici}.} \bibinfo{year}{2015}\natexlab{}.
\newblock \showarticletitle{$\{$GSMem$\}$: Data Exfiltration from $\{$Air-Gapped$\}$ Computers over $\{$GSM$\}$ Frequencies}. In \bibinfo{booktitle}{\emph{24th USENIX Security Symposium (USENIX Security 15)}}. \bibinfo{pages}{849--864}.
\newblock


\bibitem[Guri et~al\mbox{.}(2014)]%
        {AirHopper}
\bibfield{author}{\bibinfo{person}{Mordechai Guri}, \bibinfo{person}{Gabi Kedma}, \bibinfo{person}{Assaf Kachlon}, {and} \bibinfo{person}{Yuval Elovici}.} \bibinfo{year}{2014}\natexlab{}.
\newblock \showarticletitle{AirHopper: Bridging the air-gap between isolated networks and mobile phones using radio frequencies}. In \bibinfo{booktitle}{\emph{2014 9th International Conference on Malicious and Unwanted Software: The Americas (MALWARE)}}. IEEE, \bibinfo{pages}{58--67}.
\newblock


\bibitem[Guri and Monitz(2018)]%
        {LCDReloaded}
\bibfield{author}{\bibinfo{person}{Mordechai Guri} {and} \bibinfo{person}{Matan Monitz}.} \bibinfo{year}{2018}\natexlab{}.
\newblock \showarticletitle{Lcd tempest air-gap attack reloaded}. In \bibinfo{booktitle}{\emph{2018 IEEE International Conference on the Science of Electrical Engineering in Israel (ICSEE)}}. IEEE, \bibinfo{pages}{1--5}.
\newblock


\bibitem[Guri et~al\mbox{.}(2016)]%
        {USBee}
\bibfield{author}{\bibinfo{person}{Mordechai Guri}, \bibinfo{person}{Matan Monitz}, {and} \bibinfo{person}{Yuval Elovici}.} \bibinfo{year}{2016}\natexlab{}.
\newblock \showarticletitle{USBee: Air-gap covert-channel via electromagnetic emission from USB}. In \bibinfo{booktitle}{\emph{2016 14th Annual Conference on Privacy, Security and Trust (PST)}}. IEEE, \bibinfo{pages}{264--268}.
\newblock


\bibitem[Hayashi et~al\mbox{.}(2014)]%
        {hayashi2014threat}
\bibfield{author}{\bibinfo{person}{Yuichi Hayashi}, \bibinfo{person}{Naofumi Homma}, \bibinfo{person}{Mamoru Miura}, \bibinfo{person}{Takafumi Aoki}, {and} \bibinfo{person}{Hideaki Sone}.} \bibinfo{year}{2014}\natexlab{}.
\newblock \showarticletitle{A threat for tablet pcs in public space: Remote visualization of screen images using em emanation}. In \bibinfo{booktitle}{\emph{Proceedings of the 2014 ACM SIGSAC Conference on Computer and Communications Security}}. \bibinfo{pages}{954--965}.
\newblock


\bibitem[Hemming(2000)]%
        {MilitaryShielding}
\bibfield{author}{\bibinfo{person}{Leland~H. Hemming}.} \bibinfo{year}{2000}\natexlab{}.
\newblock \bibinfo{booktitle}{\emph{Architectural Electromagnetic Shielding Handbook: A Design and Specification Guide}}.
\newblock \bibinfo{publisher}{John Wiley \& Sons}.
\newblock


\bibitem[Jiang et~al\mbox{.}(2017)]%
        {Bluebee}
\bibfield{author}{\bibinfo{person}{Wenchao Jiang}, \bibinfo{person}{Zhimeng Yin}, \bibinfo{person}{Ruofeng Liu}, \bibinfo{person}{Zhijun Li}, \bibinfo{person}{Song~Min Kim}, {and} \bibinfo{person}{Tian He}.} \bibinfo{year}{2017}\natexlab{}.
\newblock \showarticletitle{Bluebee: a 10,000 x faster cross-technology communication via phy emulation}. In \bibinfo{booktitle}{\emph{Proceedings of the 15th ACM Conference on Embedded Network Sensor Systems}}. \bibinfo{pages}{1--13}.
\newblock


\bibitem[Jouhari et~al\mbox{.}(2023)]%
        {jouhari2023survey}
\bibfield{author}{\bibinfo{person}{Mohammed Jouhari}, \bibinfo{person}{Nasir Saeed}, \bibinfo{person}{Mohamed-Slim Alouini}, {and} \bibinfo{person}{El~Mehdi Amhoud}.} \bibinfo{year}{2023}\natexlab{}.
\newblock \showarticletitle{A survey on scalable LoRaWAN for massive IoT: Recent advances, potentials, and challenges}.
\newblock \bibinfo{journal}{\emph{IEEE Communications Surveys \& Tutorials}} (\bibinfo{year}{2023}).
\newblock


\bibitem[Knapp(2024)]%
        {knapp2024industrial}
\bibfield{author}{\bibinfo{person}{Eric~D Knapp}.} \bibinfo{year}{2024}\natexlab{}.
\newblock \bibinfo{booktitle}{\emph{Industrial Network Security: Securing critical infrastructure networks for smart grid, SCADA, and other Industrial Control Systems}}.
\newblock \bibinfo{publisher}{Elsevier}.
\newblock


\bibitem[Kolobe et~al\mbox{.}(2020)]%
        {kolobe2020systematic}
\bibfield{author}{\bibinfo{person}{Lone Kolobe}, \bibinfo{person}{Boyce Sigweni}, {and} \bibinfo{person}{Caspar~K Lebekwe}.} \bibinfo{year}{2020}\natexlab{}.
\newblock \showarticletitle{Systematic literature survey: Applications of LoRa communications}.
\newblock  (\bibinfo{year}{2020}).
\newblock


\bibitem[Koyun and Al~Janabi(2017)]%
        {Social1}
\bibfield{author}{\bibinfo{person}{Arif Koyun} {and} \bibinfo{person}{Ehssan Al~Janabi}.} \bibinfo{year}{2017}\natexlab{}.
\newblock \showarticletitle{Social engineering attacks}.
\newblock \bibinfo{journal}{\emph{Journal of Multidisciplinary Engineering Science and Technology (JMEST)}} \bibinfo{volume}{4}, \bibinfo{number}{6} (\bibinfo{year}{2017}), \bibinfo{pages}{7533--7538}.
\newblock


\bibitem[Kuhn and Anderson(1998)]%
        {SoftTEMPEST}
\bibfield{author}{\bibinfo{person}{Markus~G Kuhn} {and} \bibinfo{person}{Ross~J Anderson}.} \bibinfo{year}{1998}\natexlab{}.
\newblock \showarticletitle{Soft tempest: Hidden data transmission using electromagnetic emanations}. In \bibinfo{booktitle}{\emph{International Workshop on Information Hiding}}. Springer, \bibinfo{pages}{124--142}.
\newblock


\bibitem[Lai and Dai(2009)]%
        {lai2009implementation}
\bibfield{author}{\bibinfo{person}{Yeu-Pong Lai} {and} \bibinfo{person}{Ruan-Han Dai}.} \bibinfo{year}{2009}\natexlab{}.
\newblock \showarticletitle{The implementation guidance for practicing network isolation by referring to ISO-17799 standard}.
\newblock \bibinfo{journal}{\emph{Computer Standards \& Interfaces}} \bibinfo{volume}{31}, \bibinfo{number}{4} (\bibinfo{year}{2009}), \bibinfo{pages}{748--756}.
\newblock


\bibitem[Lampson(1973)]%
        {PhysicalCovertChannel}
\bibfield{author}{\bibinfo{person}{Butler~W Lampson}.} \bibinfo{year}{1973}\natexlab{}.
\newblock \showarticletitle{A note on the confinement problem}.
\newblock \bibinfo{journal}{\emph{Commun. ACM}} \bibinfo{volume}{16}, \bibinfo{number}{10} (\bibinfo{year}{1973}), \bibinfo{pages}{613--615}.
\newblock


\bibitem[Langner(2011)]%
        {Stuxnet}
\bibfield{author}{\bibinfo{person}{Ralph Langner}.} \bibinfo{year}{2011}\natexlab{}.
\newblock \showarticletitle{Stuxnet: Dissecting a cyberwarfare weapon}.
\newblock \bibinfo{journal}{\emph{IEEE Security \& Privacy}} \bibinfo{volume}{9}, \bibinfo{number}{3} (\bibinfo{year}{2011}), \bibinfo{pages}{49--51}.
\newblock


\bibitem[Larroca et~al\mbox{.}(2022)]%
        {gr-tempest}
\bibfield{author}{\bibinfo{person}{Federico Larroca}, \bibinfo{person}{Pablo Bertrand}, \bibinfo{person}{Felipe Carrau}, {and} \bibinfo{person}{Victoria Severi}.} \bibinfo{year}{2022}\natexlab{}.
\newblock \showarticletitle{gr-tempest: an open-source GNU Radio implementation of TEMPEST}. In \bibinfo{booktitle}{\emph{2022 Asian Hardware Oriented Security and Trust Symposium (AsianHOST)}}. IEEE, \bibinfo{pages}{1--6}.
\newblock


\bibitem[Lavaud et~al\mbox{.}(2021)]%
        {lavaud2021whispering}
\bibfield{author}{\bibinfo{person}{Corentin Lavaud}, \bibinfo{person}{Robin Gerzaguet}, \bibinfo{person}{Matthieu Gautier}, \bibinfo{person}{Olivier Berder}, \bibinfo{person}{Erwan Nogues}, {and} \bibinfo{person}{Stephane Molton}.} \bibinfo{year}{2021}\natexlab{}.
\newblock \showarticletitle{Whispering devices: A survey on how side-channels lead to compromised information}.
\newblock \bibinfo{journal}{\emph{Journal of Hardware and Systems Security}}  \bibinfo{volume}{5} (\bibinfo{year}{2021}), \bibinfo{pages}{143--168}.
\newblock


\bibitem[Lee et~al\mbox{.}(2022)]%
        {TMDSLeakageAnalysis}
\bibfield{author}{\bibinfo{person}{Euibum Lee}, \bibinfo{person}{Dong-Hoon Choi}, \bibinfo{person}{Taesik Nam}, {and} \bibinfo{person}{Jong-Gwan Yook}.} \bibinfo{year}{2022}\natexlab{}.
\newblock \showarticletitle{A quantitative analysis of compromising emanation from TMDS interface and possibility of sensitive information leakage}.
\newblock \bibinfo{journal}{\emph{IEEE Access}}  \bibinfo{volume}{10} (\bibinfo{year}{2022}), \bibinfo{pages}{73997--74011}.
\newblock


\bibitem[Lee et~al\mbox{.}(2015)]%
        {Hidden_pixels}
\bibfield{author}{\bibinfo{person}{Ho~Seong Lee}, \bibinfo{person}{Jong-Gwan Yook}, {and} \bibinfo{person}{Kyuhong Sim}.} \bibinfo{year}{2015}\natexlab{}.
\newblock \showarticletitle{Measurement and analysis of the electromagnetic emanations from video display interface}. In \bibinfo{booktitle}{\emph{2015 IEEE Electrical Design of Advanced Packaging and Systems Symposium (EDAPS)}}. IEEE, \bibinfo{pages}{71--73}.
\newblock


\bibitem[Lendvay(2016)]%
        {lendvay2016shadows}
\bibfield{author}{\bibinfo{person}{Ronald~L Lendvay}.} \bibinfo{year}{2016}\natexlab{}.
\newblock \emph{\bibinfo{title}{Shadows of Stuxnet: Recommendations for US policy on critical infrastructure cyber defense derived from the Stuxnet attack}}.
\newblock \bibinfo{thesistype}{Ph.\,D. Dissertation}. \bibinfo{school}{Monterey, California: Naval Postgraduate School}.
\newblock


\bibitem[Leonardi et~al\mbox{.}(2019)]%
        {leonardi2019rt}
\bibfield{author}{\bibinfo{person}{Luca Leonardi}, \bibinfo{person}{Filippo Battaglia}, {and} \bibinfo{person}{Lucia~Lo Bello}.} \bibinfo{year}{2019}\natexlab{}.
\newblock \showarticletitle{RT-LoRa: A medium access strategy to support real-time flows over LoRa-based networks for industrial IoT applications}.
\newblock \bibinfo{journal}{\emph{IEEE Internet of Things Journal}} \bibinfo{volume}{6}, \bibinfo{number}{6} (\bibinfo{year}{2019}), \bibinfo{pages}{10812--10823}.
\newblock


\bibitem[Li and Cao(2022)]%
        {li2022lora}
\bibfield{author}{\bibinfo{person}{Chenning Li} {and} \bibinfo{person}{Zhichao Cao}.} \bibinfo{year}{2022}\natexlab{}.
\newblock \showarticletitle{Lora networking techniques for large-scale and long-term iot: A down-to-top survey}.
\newblock \bibinfo{journal}{\emph{ACM Computing Surveys (CSUR)}} \bibinfo{volume}{55}, \bibinfo{number}{3} (\bibinfo{year}{2022}), \bibinfo{pages}{1--36}.
\newblock


\bibitem[Li and Chen(2020)]%
        {BLE2LoRa}
\bibfield{author}{\bibinfo{person}{Zhijun Li} {and} \bibinfo{person}{Yongrui Chen}.} \bibinfo{year}{2020}\natexlab{}.
\newblock \showarticletitle{BLE2LoRa: Cross-technology communication from bluetooth to LoRa via chirp emulation}. In \bibinfo{booktitle}{\emph{2020 17th Annual IEEE International Conference on Sensing, Communication, and Networking (SECON)}}. IEEE, \bibinfo{pages}{1--9}.
\newblock


\bibitem[Li and He(2017)]%
        {Webee}
\bibfield{author}{\bibinfo{person}{Zhijun Li} {and} \bibinfo{person}{Tian He}.} \bibinfo{year}{2017}\natexlab{}.
\newblock \showarticletitle{Webee: Physical-layer cross-technology communication via emulation}. In \bibinfo{booktitle}{\emph{Proceedings of the 23rd Annual International Conference on Mobile Computing and Networking}}. \bibinfo{pages}{2--14}.
\newblock


\bibitem[Lilygo(2024)]%
        {Lilygo}
\bibfield{author}{\bibinfo{person}{Lilygo}.} \bibinfo{year}{2024}\natexlab{}.
\newblock \bibinfo{booktitle}{\emph{Official Website [online]}}.
\newblock
\urldef\tempurl%
\url{https://www.lilygo.cc}
\showURL{%
\tempurl}


\bibitem[Lin et~al\mbox{.}(2010)]%
        {lin2010overshoot}
\bibfield{author}{\bibinfo{person}{Hui Lin}, \bibinfo{person}{Liming Wu}, \bibinfo{person}{Junxiu Liu}, {and} \bibinfo{person}{Tengteng Wen}.} \bibinfo{year}{2010}\natexlab{}.
\newblock \showarticletitle{Overshoot and undershoot control for signal generator}. In \bibinfo{booktitle}{\emph{2010 International Conference on Measuring Technology and Mechatronics Automation}}, Vol.~\bibinfo{volume}{2}. IEEE, \bibinfo{pages}{864--867}.
\newblock


\bibitem[Liu et~al\mbox{.}(2020)]%
        {PhoneTEMPEST}
\bibfield{author}{\bibinfo{person}{Zhuoran Liu}, \bibinfo{person}{Niels Samwel}, \bibinfo{person}{L{\'e}o Weissbart}, \bibinfo{person}{Zhengyu Zhao}, \bibinfo{person}{Dirk Lauret}, \bibinfo{person}{Lejla Batina}, {and} \bibinfo{person}{Martha Larson}.} \bibinfo{year}{2020}\natexlab{}.
\newblock \showarticletitle{Screen Gleaning: A Screen Reading TEMPEST Attack on Mobile Devices Exploiting an Electromagnetic Side Channel}.
\newblock \bibinfo{journal}{\emph{arXiv preprint arXiv:2011.09877}} (\bibinfo{year}{2020}).
\newblock


\bibitem[Long et~al\mbox{.}(2019)]%
        {EMeye}
\bibfield{author}{\bibinfo{person}{Yan Long}, \bibinfo{person}{Qinhong Jiang}, \bibinfo{person}{Chen Yan}, \bibinfo{person}{Tobias Alam}, \bibinfo{person}{Xiaoyu Ji}, \bibinfo{person}{Wenyuan Xu}, {and} \bibinfo{person}{Kevin Fu}.} \bibinfo{year}{2019}\natexlab{}.
\newblock \showarticletitle{EM Eye: Characterizing Electromagnetic Side-channel Eavesdropping on Embedded Cameras}.
\newblock  (\bibinfo{year}{2019}).
\newblock


\bibitem[Marinov(2014)]%
        {TEMPESTSDR}
\bibfield{author}{\bibinfo{person}{Martin Marinov}.} \bibinfo{year}{2014}\natexlab{}.
\newblock \showarticletitle{Remote video eavesdropping using a software-defined radio platform}.
\newblock \bibinfo{journal}{\emph{MS thesis, University of Cambridge}} (\bibinfo{year}{2014}).
\newblock


\bibitem[Migabo et~al\mbox{.}(2018)]%
        {NB-IoT-Protocol}
\bibfield{author}{\bibinfo{person}{Emmanuel Migabo}, \bibinfo{person}{Karim Djouani}, {and} \bibinfo{person}{Anish Kurien}.} \bibinfo{year}{2018}\natexlab{}.
\newblock \showarticletitle{A modelling approach for the narrowband IoT (NB-IoT) physical (PHY) layer performance}. In \bibinfo{booktitle}{\emph{IECON 2018-44th Annual Conference of the IEEE Industrial Electronics Society}}. IEEE, \bibinfo{pages}{5207--5214}.
\newblock


\bibitem[Organisation(2021)]%
        {ECSSecurity1}
\bibfield{author}{\bibinfo{person}{European Cyber~Security Organisation}.} \bibinfo{year}{2021}\natexlab{}.
\newblock \bibinfo{booktitle}{\emph{System security and certification considerations [online]}}.
\newblock
\urldef\tempurl%
\url{https://ecs-org.eu/ecso-uploads/2022/10/61ebc4a13b567.pdf}
\showURL{%
\tempurl}


\bibitem[Organisation(2022)]%
        {ECSSecurity2}
\bibfield{author}{\bibinfo{person}{European Cyber~Security Organisation}.} \bibinfo{year}{2022}\natexlab{}.
\newblock \bibinfo{booktitle}{\emph{ECSO Technical Paper on Internet of Things (IoT) [online]}}.
\newblock
\urldef\tempurl%
\url{https://ecs-org.eu/ecso-uploads/2023/01/ECSO_WG6_IoT-Technical_paper_final.pdf}
\showURL{%
\tempurl}


\bibitem[Paul et~al\mbox{.}(2022)]%
        {paul2022introduction}
\bibfield{author}{\bibinfo{person}{Clayton~R Paul}, \bibinfo{person}{Robert~C Scully}, {and} \bibinfo{person}{Mark~A Steffka}.} \bibinfo{year}{2022}\natexlab{}.
\newblock \bibinfo{booktitle}{\emph{Introduction to electromagnetic compatibility}}.
\newblock \bibinfo{publisher}{John Wiley \& Sons}.
\newblock


\bibitem[Peisert et~al\mbox{.}(2021)]%
        {SolarWinds}
\bibfield{author}{\bibinfo{person}{Sean Peisert}, \bibinfo{person}{Bruce Schneier}, \bibinfo{person}{Hamed Okhravi}, \bibinfo{person}{Fabio Massacci}, \bibinfo{person}{Terry Benzel}, \bibinfo{person}{Carl Landwehr}, \bibinfo{person}{Mohammad Mannan}, \bibinfo{person}{Jelena Mirkovic}, \bibinfo{person}{Atul Prakash}, {and} \bibinfo{person}{James~Bret Michael}.} \bibinfo{year}{2021}\natexlab{}.
\newblock \showarticletitle{Perspectives on the SolarWinds incident}.
\newblock \bibinfo{journal}{\emph{IEEE Security \& Privacy}} \bibinfo{volume}{19}, \bibinfo{number}{2} (\bibinfo{year}{2021}), \bibinfo{pages}{7--13}.
\newblock


\bibitem[Pub(1994)]%
        {pub1994security}
\bibfield{author}{\bibinfo{person}{FIPS Pub}.} \bibinfo{year}{1994}\natexlab{}.
\newblock \showarticletitle{Security requirements for cryptographic modules}.
\newblock \bibinfo{journal}{\emph{FIPS PUB}}  \bibinfo{volume}{140} (\bibinfo{year}{1994}), \bibinfo{pages}{140--2}.
\newblock


\bibitem[Ramya et~al\mbox{.}(2011)]%
        {ZigBee}
\bibfield{author}{\bibinfo{person}{C~Muthu Ramya}, \bibinfo{person}{Madasamy Shanmugaraj}, {and} \bibinfo{person}{R Prabakaran}.} \bibinfo{year}{2011}\natexlab{}.
\newblock \showarticletitle{Study on ZigBee technology}. In \bibinfo{booktitle}{\emph{2011 3rd international conference on electronics computer technology}}, Vol.~\bibinfo{volume}{6}. IEEE, \bibinfo{pages}{297--301}.
\newblock


\bibitem[Roberts(2006)]%
        {RFID433MHz}
\bibfield{author}{\bibinfo{person}{Chris~M Roberts}.} \bibinfo{year}{2006}\natexlab{}.
\newblock \showarticletitle{Radio frequency identification (RFID)}.
\newblock \bibinfo{journal}{\emph{Computers \& security}} \bibinfo{volume}{25}, \bibinfo{number}{1} (\bibinfo{year}{2006}), \bibinfo{pages}{18--26}.
\newblock


\bibitem[Sanmillan(2020)]%
        {Ramsay}
\bibfield{author}{\bibinfo{person}{Ignacio Sanmillan}.} \bibinfo{year}{2020}\natexlab{}.
\newblock \showarticletitle{Ramsay: A cyber-espionage toolkit tailored for air-gapped networks}.
\newblock \bibinfo{journal}{\emph{Retrieved October}}  \bibinfo{volume}{12} (\bibinfo{year}{2020}), \bibinfo{pages}{2020}.
\newblock


\bibitem[Sayakkara et~al\mbox{.}(2018)]%
        {sayakkara2018accuracy}
\bibfield{author}{\bibinfo{person}{Asanka Sayakkara}, \bibinfo{person}{Nhien-An Le-Khac}, {and} \bibinfo{person}{Mark Scanlon}.} \bibinfo{year}{2018}\natexlab{}.
\newblock \showarticletitle{Accuracy enhancement of electromagnetic side-channel attacks on computer monitors}. In \bibinfo{booktitle}{\emph{Proceedings of the 13th International Conference on Availability, Reliability and Security}}. \bibinfo{pages}{1--9}.
\newblock


\bibitem[Scarfone et~al\mbox{.}(2008)]%
        {scarfone2008technical}
\bibfield{author}{\bibinfo{person}{Karen Scarfone}, \bibinfo{person}{Murugiah Souppaya}, \bibinfo{person}{Amanda Cody}, {and} \bibinfo{person}{Angela Orebaugh}.} \bibinfo{year}{2008}\natexlab{}.
\newblock \showarticletitle{Technical guide to information security testing and assessment}.
\newblock \bibinfo{journal}{\emph{NIST Special Publication}} \bibinfo{volume}{800}, \bibinfo{number}{115} (\bibinfo{year}{2008}), \bibinfo{pages}{2--25}.
\newblock


\bibitem[Shen et~al\mbox{.}(2021)]%
        {EMLoRa}
\bibfield{author}{\bibinfo{person}{Cheng Shen}, \bibinfo{person}{Tian Liu}, \bibinfo{person}{Jun Huang}, {and} \bibinfo{person}{Rui Tan}.} \bibinfo{year}{2021}\natexlab{}.
\newblock \showarticletitle{When LoRa meets EMR: Electromagnetic covert channels can be super resilient}. In \bibinfo{booktitle}{\emph{2021 IEEE Symposium on Security and Privacy (SP)}}. IEEE, \bibinfo{pages}{1304--1317}.
\newblock


\bibitem[Shi et~al\mbox{.}(2019)]%
        {Lorabee}
\bibfield{author}{\bibinfo{person}{Junyang Shi}, \bibinfo{person}{Di Mu}, {and} \bibinfo{person}{Mo Sha}.} \bibinfo{year}{2019}\natexlab{}.
\newblock \showarticletitle{Lorabee: Cross-technology communication from lora to zigbee via payload encoding}. In \bibinfo{booktitle}{\emph{2019 IEEE 27th International Conference on Network Protocols (ICNP)}}. IEEE, \bibinfo{pages}{1--11}.
\newblock


\bibitem[Siafarikas and Volakis(2020)]%
        {Direct-RF}
\bibfield{author}{\bibinfo{person}{Dimitrios Siafarikas} {and} \bibinfo{person}{John~L Volakis}.} \bibinfo{year}{2020}\natexlab{}.
\newblock \showarticletitle{Toward direct RF sampling: Implications for digital communications}.
\newblock \bibinfo{journal}{\emph{IEEE Microwave Magazine}} \bibinfo{volume}{21}, \bibinfo{number}{9} (\bibinfo{year}{2020}), \bibinfo{pages}{43--52}.
\newblock


\bibitem[Stouffer et~al\mbox{.}(2011)]%
        {stouffer2011guide}
\bibfield{author}{\bibinfo{person}{Keith Stouffer}, \bibinfo{person}{Joe Falco}, \bibinfo{person}{Karen Scarfone}, {et~al\mbox{.}}} \bibinfo{year}{2011}\natexlab{}.
\newblock \showarticletitle{Guide to industrial control systems (ICS) security}.
\newblock \bibinfo{journal}{\emph{NIST special publication}} \bibinfo{volume}{800}, \bibinfo{number}{82} (\bibinfo{year}{2011}), \bibinfo{pages}{16--16}.
\newblock


\bibitem[Thiele(2001)]%
        {TEMPESTForEliza}
\bibfield{author}{\bibinfo{person}{Erik Thiele}.} \bibinfo{year}{2001}\natexlab{}.
\newblock \showarticletitle{Tempest for Eliza}.
\newblock \bibinfo{journal}{\emph{[Online] http://www.erikyyy.de/tempest}} (\bibinfo{year}{2001}).
\newblock


\bibitem[to~1020MHz Long Range Low Power~Transceiver(2024)]%
        {SX1262}
\bibfield{author}{\bibinfo{person}{Semtech LoRa Connect™~137MHz to 1020MHz Long Range Low Power~Transceiver}.} \bibinfo{year}{2024}\natexlab{}.
\newblock \bibinfo{booktitle}{\emph{SX1262 Official Website [online]}}.
\newblock
\urldef\tempurl%
\url{https://www.semtech.com/products/wireless-rf/lora-connect/sx1262}
\showURL{%
\tempurl}


\bibitem[Tong et~al\mbox{.}(2022)]%
        {L2X}
\bibfield{author}{\bibinfo{person}{Shuai Tong}, \bibinfo{person}{Yangliang He}, \bibinfo{person}{Yunhao Liu}, {and} \bibinfo{person}{Jiliang Wang}.} \bibinfo{year}{2022}\natexlab{}.
\newblock \showarticletitle{De-spreading over the air: long-range ctc for diverse receivers with lora}. In \bibinfo{booktitle}{\emph{Proceedings of the 28th Annual International Conference on Mobile Computing and Networking}}. \bibinfo{pages}{42--54}.
\newblock


\bibitem[Van~Eck(1985)]%
        {Van1985}
\bibfield{author}{\bibinfo{person}{Wim Van~Eck}.} \bibinfo{year}{1985}\natexlab{}.
\newblock \showarticletitle{Electromagnetic radiation from video display units: An eavesdropping risk?}
\newblock \bibinfo{journal}{\emph{Computers \& Security}} \bibinfo{volume}{4}, \bibinfo{number}{4} (\bibinfo{year}{1985}), \bibinfo{pages}{269--286}.
\newblock


\bibitem[Voigt and Von~dem Bussche(2017)]%
        {GDPR}
\bibfield{author}{\bibinfo{person}{Paul Voigt} {and} \bibinfo{person}{Axel Von~dem Bussche}.} \bibinfo{year}{2017}\natexlab{}.
\newblock \showarticletitle{The eu general data protection regulation (gdpr)}.
\newblock \bibinfo{journal}{\emph{A Practical Guide, 1st Ed., Cham: Springer International Publishing}} \bibinfo{volume}{10}, \bibinfo{number}{3152676} (\bibinfo{year}{2017}), \bibinfo{pages}{10--5555}.
\newblock


\bibitem[Wang et~al\mbox{.}(2020)]%
        {Social2}
\bibfield{author}{\bibinfo{person}{Zuoguang Wang}, \bibinfo{person}{Limin Sun}, {and} \bibinfo{person}{Hongsong Zhu}.} \bibinfo{year}{2020}\natexlab{}.
\newblock \showarticletitle{Defining social engineering in cybersecurity}.
\newblock \bibinfo{journal}{\emph{IEEE Access}}  \bibinfo{volume}{8} (\bibinfo{year}{2020}), \bibinfo{pages}{85094--85115}.
\newblock


\bibitem[WaveShare(2024)]%
        {WaveShare}
\bibfield{author}{\bibinfo{person}{WaveShare}.} \bibinfo{year}{2024}\natexlab{}.
\newblock \bibinfo{booktitle}{\emph{Official Website [online]}}.
\newblock
\urldef\tempurl%
\url{https://www.waveshare.com/}
\showURL{%
\tempurl}


\bibitem[{Wikipedia}(2024)]%
        {enwiki:1073264081}
\bibfield{author}{\bibinfo{person}{{Wikipedia}}.} \bibinfo{year}{2024}\natexlab{}.
\newblock \bibinfo{title}{Tempest (codename) --- {Wikipedia}{,} The Free Encyclopedia}.
\newblock
\urldef\tempurl%
\url{https://en.wikipedia.org/wiki/Tempest_(codename)}
\showURL{%
\tempurl}


\bibitem[Xia et~al\mbox{.}(2022)]%
        {WiRa}
\bibfield{author}{\bibinfo{person}{Dan Xia}, \bibinfo{person}{Xiaolong Zheng}, \bibinfo{person}{Fu Yu}, \bibinfo{person}{Liang Liu}, {and} \bibinfo{person}{Huadong Ma}.} \bibinfo{year}{2022}\natexlab{}.
\newblock \showarticletitle{WiRa: Enabling cross-technology communication from WiFi to LoRa with IEEE 802.11 ax}. In \bibinfo{booktitle}{\emph{Proceedings of IEEE INFOCOM}}.
\newblock


\bibitem[Xia et~al\mbox{.}(2023)]%
        {xia2023xcopy}
\bibfield{author}{\bibinfo{person}{Xianjin Xia}, \bibinfo{person}{Qianwu Chen}, \bibinfo{person}{Ningning Hou}, \bibinfo{person}{Yuanqing Zheng}, {and} \bibinfo{person}{Mo Li}.} \bibinfo{year}{2023}\natexlab{}.
\newblock \showarticletitle{XCopy: Boosting Weak Links for Reliable LoRa Communication}. In \bibinfo{booktitle}{\emph{Proceedings of the 29th Annual International Conference on Mobile Computing and Networking}}. \bibinfo{pages}{1--15}.
\newblock


\bibitem[Yang and Zheng(2023)]%
        {yang2023aquahelper}
\bibfield{author}{\bibinfo{person}{Qiang Yang} {and} \bibinfo{person}{Yuanqing Zheng}.} \bibinfo{year}{2023}\natexlab{}.
\newblock \showarticletitle{AquaHelper: Underwater sos transmission and detection in swimming pools}. In \bibinfo{booktitle}{\emph{Proceedings of the 21st ACM Conference on Embedded Networked Sensor Systems}}. \bibinfo{pages}{294--307}.
\newblock


\bibitem[Yubo et~al\mbox{.}(2013)]%
        {ZIMO}
\bibfield{author}{\bibinfo{person}{Yan Yubo}, \bibinfo{person}{Yang Panlong}, \bibinfo{person}{Li Xiangyang}, \bibinfo{person}{Tao Yue}, \bibinfo{person}{Zhang Lan}, {and} \bibinfo{person}{You Lizhao}.} \bibinfo{year}{2013}\natexlab{}.
\newblock \showarticletitle{Zimo: Building cross-technology mimo to harmonize zigbee smog with wifi flash without intervention}. In \bibinfo{booktitle}{\emph{Proceedings of the 19th annual international conference on Mobile computing \& networking}}. \bibinfo{pages}{465--476}.
\newblock


\bibitem[Zhan et~al\mbox{.}(2020)]%
        {Bitjabber}
\bibfield{author}{\bibinfo{person}{Zihao Zhan}, \bibinfo{person}{Zhenkai Zhang}, {and} \bibinfo{person}{Xenofon Koutsoukos}.} \bibinfo{year}{2020}\natexlab{}.
\newblock \showarticletitle{Bitjabber: The world’s fastest electromagnetic covert channel}. In \bibinfo{booktitle}{\emph{2020 IEEE International Symposium on Hardware Oriented Security and Trust (HOST)}}. IEEE, \bibinfo{pages}{35--45}.
\newblock


\bibitem[ZYGIEREWICZ(2020)]%
        {anna2020directive}
\bibfield{author}{\bibinfo{person}{Anna ZYGIEREWICZ}.} \bibinfo{year}{2020}\natexlab{}.
\newblock \showarticletitle{Directive on security of network and information systems (NIS Directive)}.
\newblock  (\bibinfo{year}{2020}).
\newblock


\end{thebibliography}

\end{document}